\documentclass[usenatbib]{mn2e}
\usepackage{fixltx2e,longtable,ifpdf,afterpage}
\ifpdf
  \usepackage{aeguill}
  \usepackage[pdftex]{graphicx,color}
\else
  \usepackage[dvips]{graphicx,color}
\fi

\newcommand\2{{\sc{ii}}}

\newcommand{\h}{{$H$}}
\newcommand{\ks}{$K_{\rm s}$}

\title[VLT-MAD observations of the core of 30~Doradus]{VLT-MAD observations of the core of 30~Doradus}

\author[Campbell et al.]
{M.~A.~Campbell$^{1}$, C.~J.~Evans$^{2,1}$, A.~D.~Mackey$^{1}$, M.~Gieles$^{3}$, J.~Alves$^{4}$, 
\newauthor J.~Ascenso$^{5}$, N.~Bastian$^{6}$ \& A.~J.~Longmore$^{2}$
\\
$^{1}$Institute for Astronomy, The University of Edinburgh, Royal Observatory, Blackford Hill, Edinburgh EH9 3HJ, UK;\\
$^{2}$UK Astronomy Technology Centre, Royal Observatory, Blackford Hill, Edinburgh, EH9 3HJ, UK; \\
$^{3}$European Southern Observatory, Casilla 19001, Santiago 19, Chile;\\
$^{4}$Calar Alto Observatory-Centro Astron\'{o}mico Hispano-Alem\'{a}n, C/ Jes\'{u}s Durb\'{a}n Rem\'{o}n 2-2, 04004 Almeria, Spain;\\
$^{5}$Harvard-Smithsonian Center for Astrophysics, 60 Garden Street, Cambridge, MA 02138, USA; \\
$^{6}$Institute of Astronomy, University of Cambridge, Madingley Road, Cambridge, CB3 0HA, UK; \\
}
\date{Received:}

\begin{document}
\maketitle

\begin{abstract}
We present \h- and \ks-band imaging of three fields at the centre of
30~Doradus in the Large Magellanic Cloud, obtained as part of the
Science Demonstration programme with the Multi-conjugate Adaptive
optics Demonstrator (MAD) at the Very Large Telescope.
Strehl ratios of 15-30\% were achieved in the \ks-band, 
yielding near-infrared images of this dense and complex region at 
unprecedented angular resolution at these wavelengths.

The MAD data are used to construct a near-infrared luminosity profile
for R136, the cluster at the core of 30~Dor.  Using cluster profiles
of the form used by \citeauthor{eff}, we find the surface brightness
can be fit by a relatively shallow power-law function
($\gamma$\,$\sim$\,1.5-1.7) over the full extent of the MAD data,
which extends to a radius of $\sim$40$''$ ($\sim$10\,pc).  We do not
see compelling evidence for a break in the luminosity profile as seen
in optical data in the literature, arguing that cluster asymmetries
are the dominant source, although extinction effects and stars from
nearby triggered star-formation likely also contribute.  These results
highlight the need to consider cluster asymmetries and multiple
spatial components in interpretation of the luminosity profiles of
distant unresolved clusters.

We also investigate seven candidate young stellar objects reported by
Gruendl \& Chu from {\em Spitzer} observations, six of which have
apparent counterparts in the MAD images. The most interesting of
these (GC09: 053839.24~$-$690552.3) appears related to a striking bow-shock--like
feature, orientated away from both R136 and the Wolf-Rayet star
Brey~75, at distances of 19\farcs5 and 8$''$ (4.7 and 1.9\,pc in
projection), respectively.

\end{abstract}


\section{Introduction}\label{intro} 

30~Doradus in the Large Magellanic Cloud (LMC) is the largest
star-forming region in the Local Group.  As an archetypal,
small-scale `starburst', 30~Dor provides us with an excellent laboratory 
in which to study star formation and stellar evolution, while also giving us
potential insights into the nature of distant super-star-clusters for
which we only have integrated properties.  Extensive ground-based
optical imaging and spectroscopy has been used to study the initial
mass function (IMF), reddening, star-formation history, stellar
content and kinematics in 30~Dor \citep[e.g.][]{m85,p93,pg93,wb97,s98,b01}.
Meanwhile, near-IR observations with the {\em Hubble Space Telescope
(HST)} have provided evidence of triggered star-formation, showing the region to
be a two-stage starburst \citep{wal99,wal02}.

At the centre of 30~Dor is the dense star cluster R136, with stellar
ages for the most massive stars in the range of 1-2\,Myr \citep{mh98} and a total stellar mass in
the range of $\sim$0.35-1$\times10^{5}$\,M$_\odot$
\citep[e.g.][]{mg03,ng07,a09}, depending on the low-mass form of the mass function, 
putting it on a par with some of the clusters found in starburst and interacting galaxies such as M51, M82
and the Antennae. However, the core of R136 is too dense for
traditional (seeing-limited) ground-based techniques.

Only with the arrival of {\em HST} was R136 resolved in optical and UV
images \citep{c92,dm93,h95,h97,s00}, with follow-up spectroscopy revealing a
hitherto unprecedented concentration of the earliest O-type stars
\citep{mh98}.  Star counts from the optical {\em HST} images revealed
that the luminosity profile of R136 appears to be best described by
two components, with a break at 10$''$ \citep{mg03}.  New results from
{\em F160W} imaging (equivalent to \h-band) with the {\em HST} Near
Infrared Camera and Multi-Object Spectrometer (NICMOS) fit the profile
with a single component \citep{a09}, but only in
the inner 2\,pc (8\farcs25). Whether the second component seen in the
optical data is a manifestation of the `excess light' predicted to
originate from rapid gas removal in the early stages of cluster
evolution \citep{bg06} remains an open question as there is significant and
variable extinction across the cluster.  

As part of the technology development plan towards the European
Extremely Large Telescope (E-ELT), the Multi-conjugate Adaptive optics
Demonstrator \citep[MAD,][]{em07} was developed as a visiting
instrument for the Very Large Telescope (VLT).  The offer of Science
Demonstration (SD) observations with MAD presented the perfect
opportunity to obtain imaging of R136 at unprecedented angular
resolution at near-IR wavelengths.  In particular, MAD has the power
to penetrate the gas and dust more successfully than in the optical
{\em HST} images, at comparable angular resolution, to provide
empirical constraints on the outer component on the luminosity
profile.  Determining if R136 is an expanding group or a
dynamically-stable star cluster would serve as an important ingredient
in the debate on the importance of `infant mortality' of young
clusters \citep[e.g.][]{ll03,bg05,fcw05,gb06,bk07,bggt08}.

Here we present
the MAD SD observations, which deliver a cleaner point spread function
(PSF) than NICMOS, at finer angular resolution, and over a larger total field.  The instrumental
performance of MAD is discussed in Section~\ref{obsdata}, with the
astrometric and photometric calibration of the data detailed in
Sections~\ref{astrometry} and \ref{photometry}, respectively.
Following discussion of seven of the candidate young stellar objects (YSOs)
reported by \citet[][hereafter GC09]{g09} that lie within the MAD
fields (Section~\ref{ysos}), we then investigate the radial luminosity
profile of R136 via a combination of star counts and integrated-light
measurements (Section~\ref{profiles}), discussing its implications in
the context of cluster formation and observations of distant
unresolved clusters.

\section{Observations \& Data Reduction}\label{obsdata}
MAD employs three Shack-Hartmann wavefront sensors to observe three
natural guide stars (NGS) across a 2$'$ circular field, thereby
allowing tomography of the atmospheric turbulence.  The turbulence is
then corrected using two deformable mirrors (operating at $\sim$400
Hz), one conjugated to the ground-layer (i.e. 0\,km), the second
conjugated to 8.5\,km above the telescope.

The high resolution, near-IR camera used with MAD is the CAmera for
Multi Conjugate Adaptive Optics \citep[CAMCAO,][]{camcao}, which
operates over the {\em J}, \h, and \ks\/ bands, with critical (2
pixel) sampling of the diffraction-limited PSF
at 2.2$\mu$m.  The detector is a 2k$\times$2k Hawaii-2 HgCdTe array,
with a pixel scale of 0\farcs028/pixel, giving a field-of-view of 57$''$x57$''$.
A useful feature is that the camera can be moved within the 2$'$ field
without requiring positional offsets of the telescope, meaning that
the adaptive optics (AO) loop can remain closed.

\begin{figure*}
\begin{center}
\includegraphics[scale=0.6]{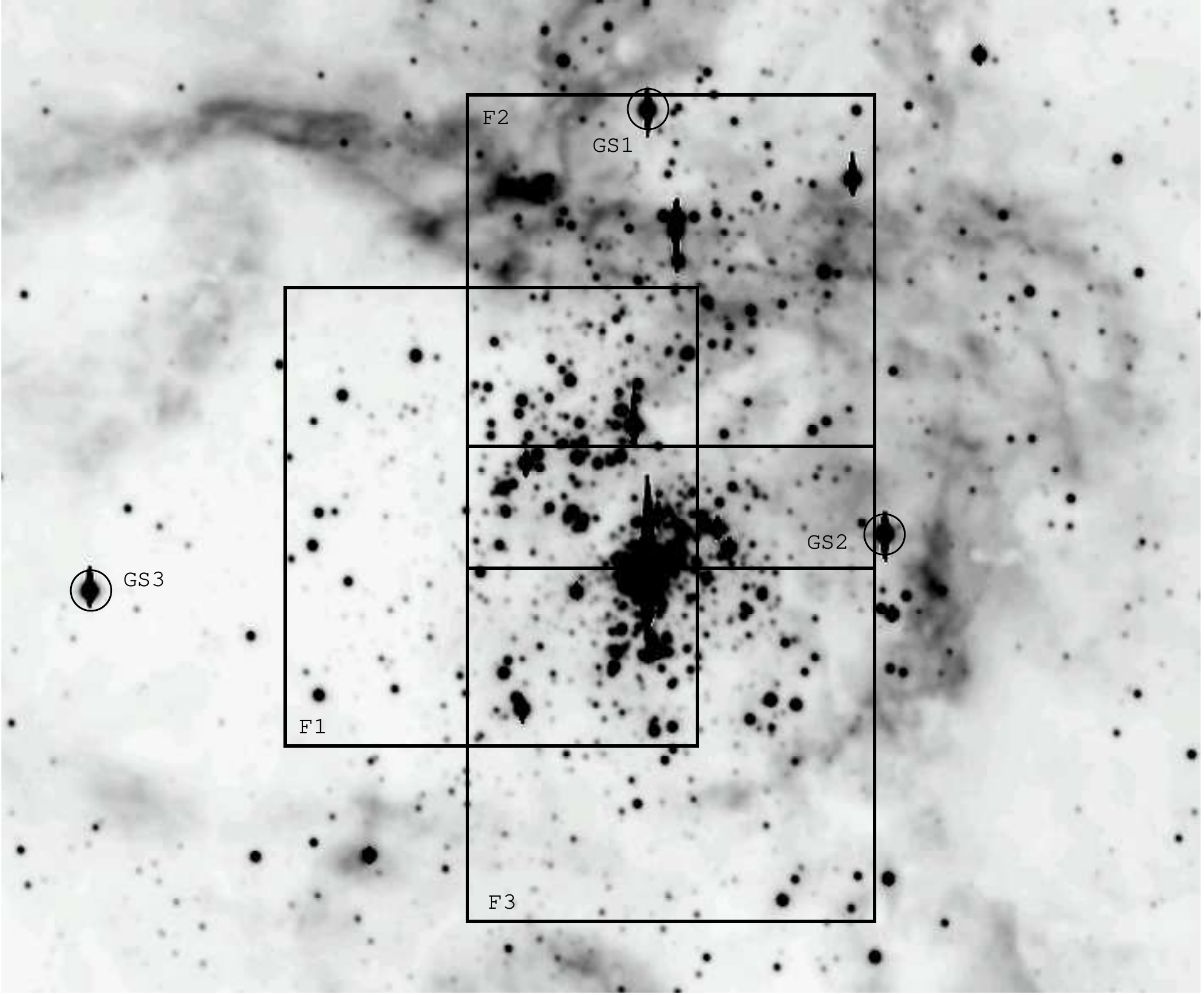}
\caption{$V$-band image of the central part of 30~Dor (approx 3\farcm0
by 2\farcm5) from the Wide-Field Imager (WFI) on the ESO/Max Planck Gesellschaft 2.2-m telescope (from observations
by J.~Alves, B.~Vandame \& Y.~Beletsky).  North is towards the top of
the figure, east towards the left.  The reference stars used for the
MCAO correction are shown (GS1, 2 \& 3), together with the spatial
extent of the dithered \h-band observations of the three fields.}\label{obs}

\includegraphics[scale=0.55]{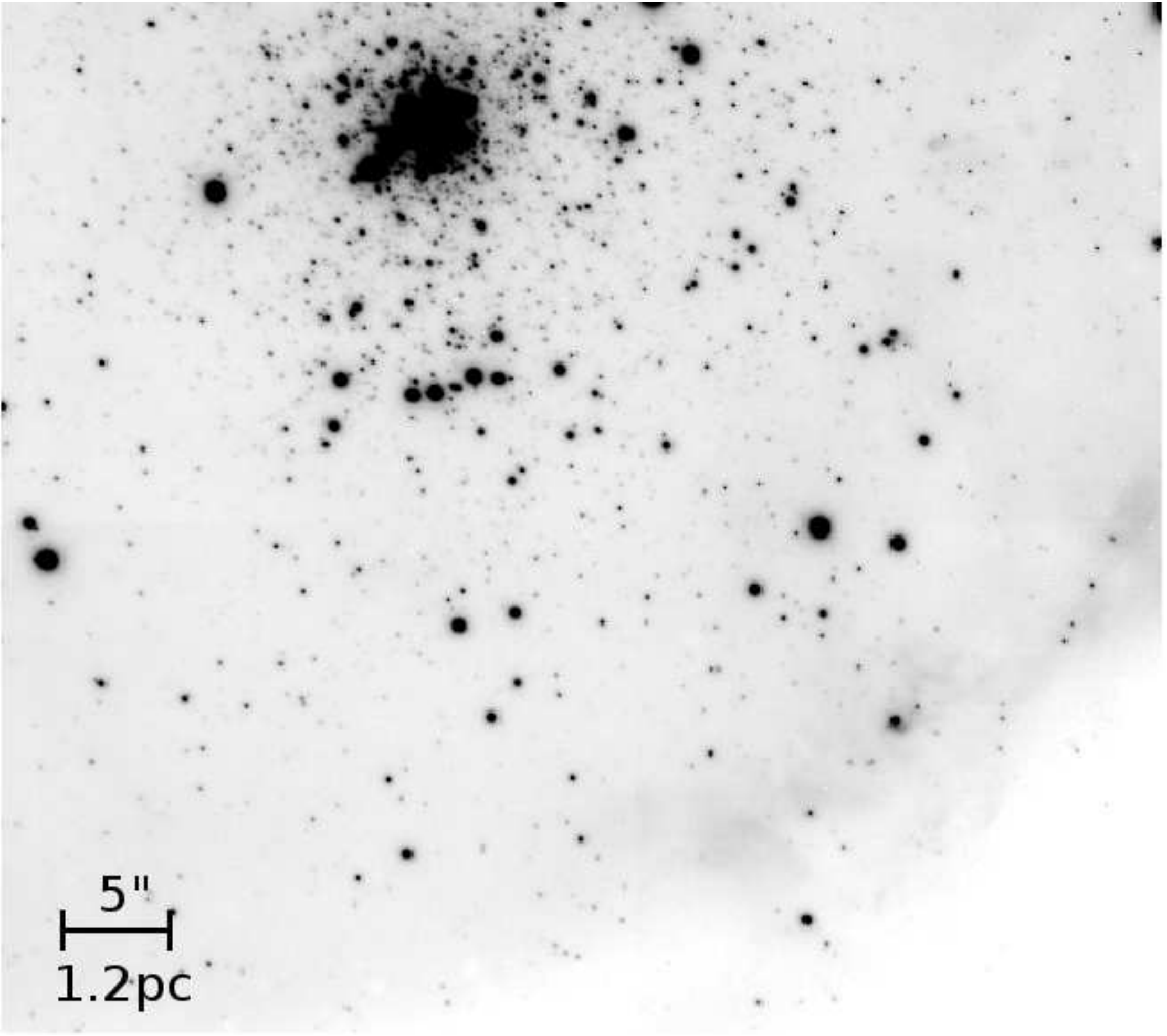}\\
\caption{MAD \ks-band image of Field~3.  Compared to the overlay on 
Figure~\ref{obs} the dithered regions are cropped, with a final field-of-view
of $\sim$52$''$\,$\times$\,46$''$.}\label{kp3}
\end{center}
\end{figure*}

The MAD data presented here were obtained from one of twelve SD
programmes, observed in November 2007 and January
2008.  \h- and \ks-band images were obtained for three pointings in
the inner region of 30~Dor, as shown in Figure~\ref{obs}.  The
central co-ordinates for the CAMCAO observations of Field 1 were
$\alpha$\,$=$\,05$^{\rm h}$38$^{\rm m}$46.5$^{\rm s}$, $\delta$\,$=$\,$-$69$^\circ$05$'$52$''$ (J2000.0).  Fields 2 and 3 were offset from
this first pointing by approximately $-$25$''$ in right ascension and $\pm$25$''$ in
declination.  The three NGS used for wavefront sensing are also shown
in Figure~\ref{obs}; these are Parker \#952 (GS1), \#499 (GS2), and
\#1788 (GS3) with {\em V}\,=\,12.0, 11.9, and 12.0, respectively \citep{p93}.
The combined \ks-band image of Field~3 is shown in Figure~\ref{kp3}.

The observations are summarised in Table~\ref{obsinfo}.  The detector
integration time (DIT) for all of the observations was 2\,s, with 30
integrations (NDIT) for each exposure.  Batches of three (\h) and six
(\ks) object and sky frames were interleaved in an A-B-A-B-A-B-A
pattern, yielding total exposures of 12\,min for each field in the
\h\/ band, and 24\,min in \ks.  Each science exposure within each
batch was dithered by 5$''$ -- although this reduces the
effective area of the final combined images, it was intended to
minimise the impact of bad pixels and cosmetic features from the
array.  Given the vast spatial extent of 30~Dor, the sky offsets were
somewhat large ($+$12s of right ascension, $+$13$'$ in declination) to
ensure they were uncontaminated by nebulosity.  Observations were
halted mid-way through the \ks\/ exposures for Field 1 on 2008 January 7
due to bad weather, but were completed the following night.  Note that
the LMC never rises above an altitude of approx. 45$^\circ$ as viewed
from Paranal, i.e. the {\em minimum} zenith distance of the
observations was $\sim$45$^\circ$, providing a good test of the MAD
performances at moderately large airmass.  Indeed, the airmass of the
observations ranged from 1.4 to 1.6.  The range of seeing values for
each field, as measured by the Differential Image Motion Monitor
(DIMM) at Paranal, are given in Table~\ref{obsinfo}.

\begin{table*}
\begin{center}
\caption{Summary of the VLT-MAD observations in 30~Doradus.  The total exposure times quoted are
for the final combined images.}\label{obsinfo}
\begin{tabular}{lclcccc}
\hline
Pointing & Band & Date & Total Exp.  & DIMM range & Image FWHM & $<$FWHM$>$ \\
& & & [min] & [$''$] & [$''$] & [$''$] \\
\hline
Field 1 &  \ks  & 2008/01/07 \& 08 & 22 & 0.4-1.8 & 0.10-0.13 & 0.11 \\
Field 2 &  \ks  & 2008/01/07           & 24 & 0.5-1.1 & 0.08-0.10 & 0.09 \\
Field 3 &  \ks  & 2007/11/27           & 23 & 0.6-1.0 & 0.10-0.20 & 0.14 \\
\hline
Field 1 & \h & 2008/01/08           & 12 & 0.3-0.6 & 0.10-0.12 & 0.11 \\
Field 2 & \h & 2008/01/08           & 12 & 0.9-1.1 & 0.08-0.11 & 0.09 \\
Field 3 & \h & 2008/01/08           & 11 & 0.6-1.6 & 0.08-0.15 & 0.12 \\
\hline
\end{tabular}
\end{center}
\end{table*}

The MAD data were reduced with standard {\sc iraf} routines, using
calibration frames from the SD runs to correct for the dark current,
to flat-field all of the object and sky exposures, and to reject bad
pixels and cosmic rays.  Median sky frames were created for each batch
of science observations using the sky frames observed immediately
before and/or after, although (in general) the sky background did not
vary strongly over each sequence of observations. The sky-subtracted
frames were then aligned with each other and combined.  At this stage
we omitted one or two images with significantly poorer image quality
in some fields, hence the final exposure times in
Table~\ref{obsinfo}. Note that no objects were saturated in any of the
individual science frames.

\subsection{Image Quality \& Performance Analysis}

\begin{figure*}
\begin{center}
\includegraphics[height=5.3cm]{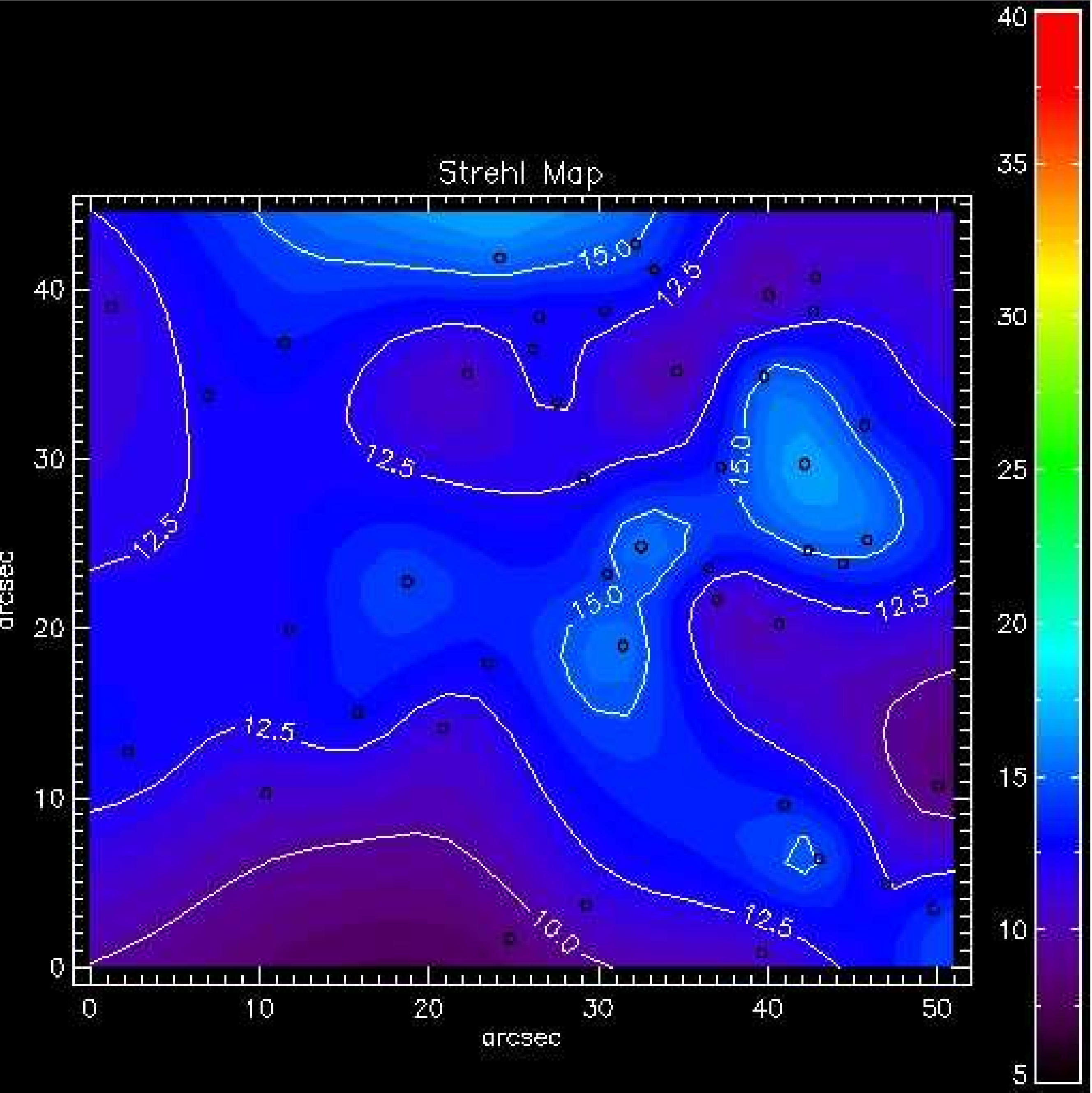}\hspace{0.1cm}\includegraphics[height=5.3cm]{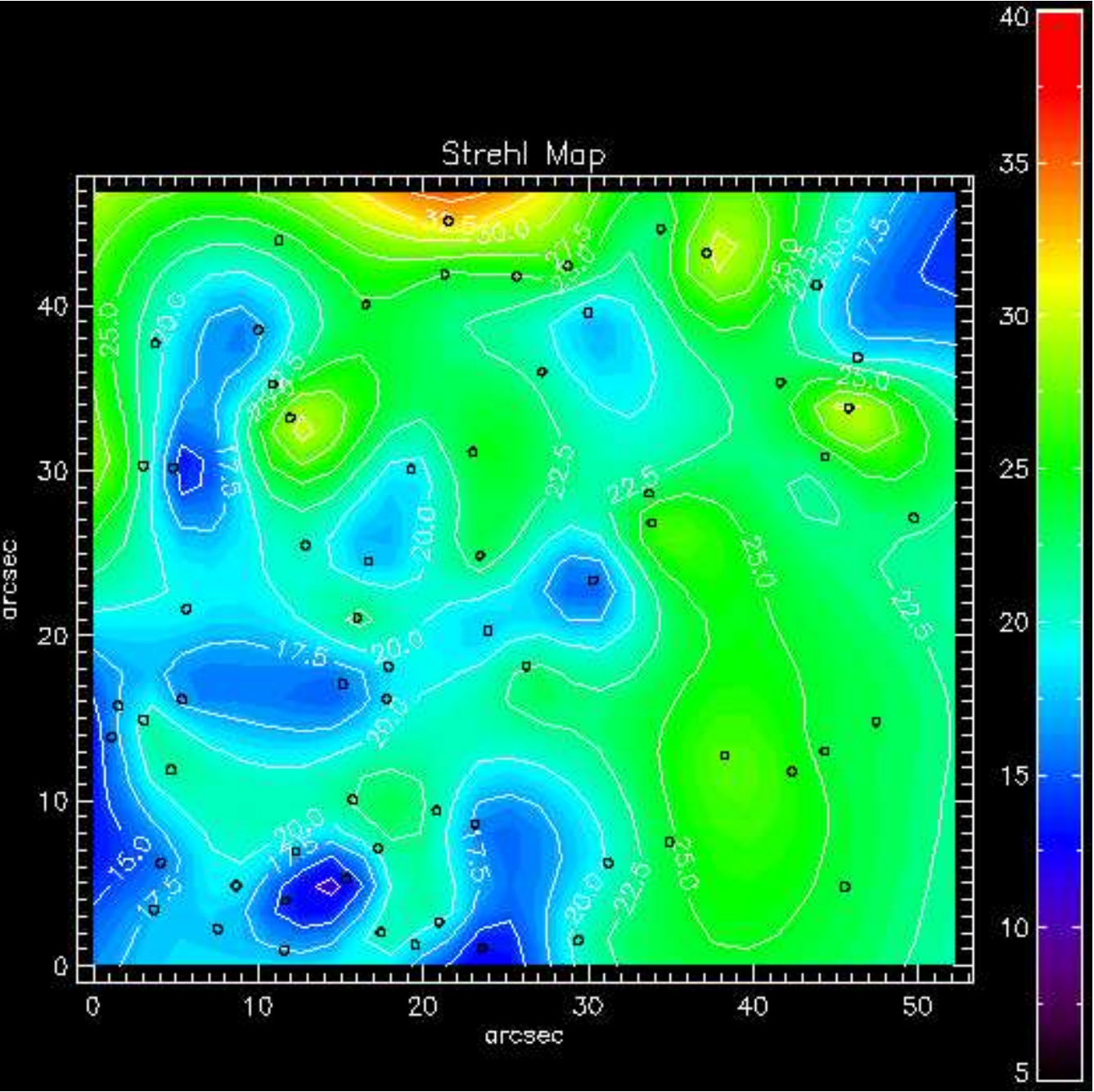}\hspace{0.1cm}\includegraphics[height=5.3cm]{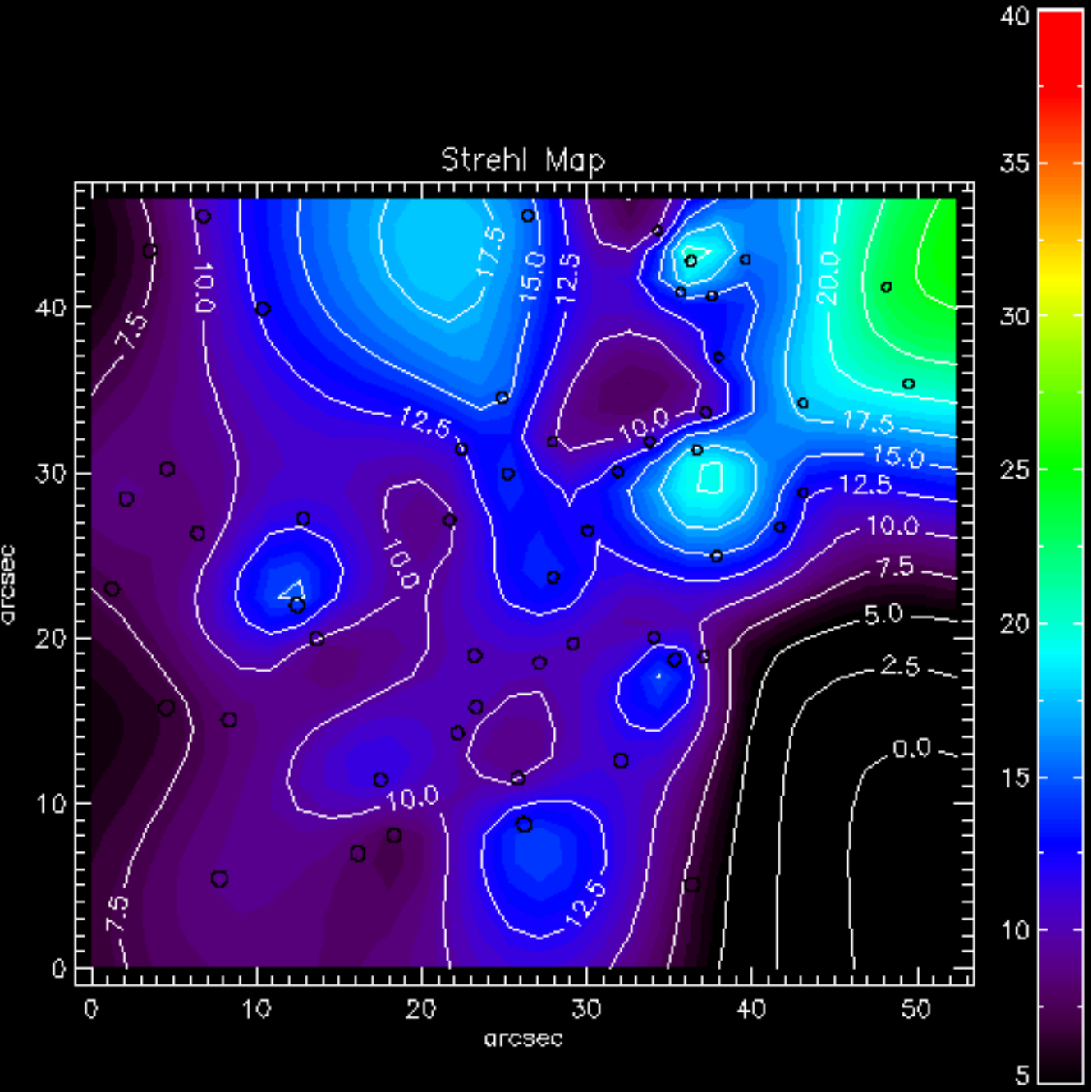}
\caption{\ks-band Strehl (\%) maps for Fields~1, 2 and 3 (left to right, respectively).  North is at the top, with east at the left.}\label{strehl}
\end{center}
\end{figure*}

Moderately bright stars were used to investigate the image quality 
in the three fields.  The range and mean of the full-width
half maxima (FWHM) obtained from 32 stars evenly distributed across
each co-added image is summarised in Table~\ref{obsinfo}.
Maps of the Strehl ratio in the \ks-band images
are shown in Figure~\ref{strehl}.  More detailed Strehl maps 
(showing the relative positions of the NGS) and FWHM maps in both
photometric bands were presented by \citet{cspie}.

The best Strehl ratio achieved in the \ks-band (25-30\%) was in the regions closest to
the NGS (as one would expect) in Fields~2 and 3.  The performance in
Field~2 is particularly good, with an average FWHM of 0\farcs09 in
both the \h\/ and \ks\/ images (compared to diffraction limits
[$\lambda$/D] of approx 0\farcs04 and 0\farcs06, respectively).
Although the Strehl is lower in Field~1, this was in the best position
with respect to all three NGS and the correction is very uniform
across the combined $\sim$50$''$x50$''$ image. This is in strong
contrast to Field 3 in which the performance is very good in one
corner, but then steeply declines away from the NGS -- more in keeping
with `classical' AO observations.  In general, the performances are
comparable with those found from other MAD observations
\citep[e.g.][]{em07,bouy08}.

\section{ Astrometric Calibration}\label{astrometry}

Astrometric calibration of each field was undertaken using the
\citet{s98} catalogue, recently recalibrated by Brian Skiff\footnote{
http://cdsarc.u-strasbg.fr/viz-bin/Cat?J/A+A/341/9}.  These positions
are dependent on the precision of the 2MASS/UCAC2 positions, combined
with the plate-scale/seeing of the original New Technology Telescope
(NTT) observations.  The quoted precision on the new Skiff astrometry
is $\sim$0\farcs1 (cf. the 0\farcs028/pixel delivered by MAD).  Visual
matches of stars between the Skiff catalogue and the MAD fields were
used to define $\sim$40 well distributed astrometric standards in
Fields~1 and 2.  The astrometric calibration of Field~3, in which
there are greater PSF variations, was achieved using 60
visually-matched stars.

\section{Photometric calibration}\label{photometry}

Photometry was performed using standard {\sc daophot} routines in {\sc
iraf} on the combined images.  Due to the variations in the PSF
delivered by the AO system across the field, PSF-fitting photometry
was used, with the model PSF ({\sc penny}{\small 2}) allowed to vary
quadratically across the fields.  Roughly 30 bright, isolated 
stars were used to create a model PSF for each combined image, with 
neighbouring stars subtracted if they were contaminating the selected
star. A fitting radius of (2$\times$FWHM)$-$1 pixels
was found to provide the best fits.
Objects with 1.5 $<$ sharpness $< -$1.5, or with $\chi <$ 4 
(where $\chi$ is the root-mean-square of the residuals that remain)
were rejected from the final catalogues to reject non-stellar objects, 
residual cosmic rays etc., as were objects with instrumental magnitude 
errors greater than 0.25$^{\rm m}$.

\subsection{Zero-point Calibrations}

\subsubsection{MAD vs. 2MASS}

Sources from the Two Micron All Sky Survey \citep[2MASS,][]{2mass}
were identified in the MAD frames for photometric calibration.
However, at the excellent angular resolution delivered by MAD many of
the 2MASS `stars' are resolved into asymmetric sources or multiple
components \citep[e.g. Fig.~3,][]{mob08} .  As such, only isolated,
apparently single stars with 2MASS photometric qualities of either
`A', `B', or `C' in the relevant band were considered for calibration.
This limited the number of 2MASS stars in each MAD field to eight or
fewer for calibration.

Photometric zero points (ZPs) were determined (principally using
`A'-rated 2MASS stars), with stars at the bright/faint ends of the
MAD images omitted to exclude strongly-exposed stars in the MAD
frames and poorly-exposed stars in 2MASS.  The resulting ZPs
are summarised in Table~\ref{zero_points}; we are somewhat 
at the mercy of small number statistics, not to mention the issues
of stellar density and complex nebular emission in and around R136.

\begin{table*}
\begin{center}
\caption{Photometric zero-points obtained from the combined image method.
Results are given for the \ks-band images using 2MASS sources within 
the MAD frames, and using the HAWK-I commissioning data, in which the result 
for Field~2 was bootstrapped using overlapping stars with Field~1.  The \h-band
results were obtained using 2MASS sources for Fields~1 and 3, with Field~2 again
calibrated using overlapping stars with Field~1.}\label{zero_points}
\begin{tabular}{llccc}
\hline
 Band & Calibration & Field 1 & Field 2 & Field 3 \\
\hline
\ks & 2MASS  &  26.78 $\pm$ 0.13 & 26.61 $\pm$ 0.45 & 26.95 $\pm$ 0.25 \\
\ks & HAWK-I &  26.69 $\pm$ 0.08 & 27.04 $\pm$ 0.09 & 27.07 $\pm$ 0.11 \\
\hline
\h  & 2MASS &  27.09 $\pm$ 0.14 & 27.27 $\pm$ 0.15 & 26.93 $\pm$ 0.08 \\
\hline
\end{tabular}
\end{center}
\end{table*}

\subsubsection {MAD vs. HAWK-I}\label{hawki}

To investigate the quality of the calibrations we employed \ks-band
commissioning data from the VLT High Acuity Wide-field K-band Imager
\citep[HAWK-I,][]{p04,c06}.  HAWK-I is a wide-field, near-IR camera
with four 2k x 2k Hawaii-2 detectors, covering a 7\farcm5 square field
at a pixel scale of 0\farcs106. Commissioning images of 30~Dor were
obtained on 2007 October 3, using $Y$, $J$, \ks\/ and Br$\gamma$
filters. The HAWK-I exposures were dithered around the centre of 30~Dor to
build-up a complete mosaic for the region\footnote{See p.\,22 of the
December 2007 edition of The Messenger.}.  

\begin{figure}
\begin{center}
\hspace{-0.85cm}\includegraphics[height=6cm]{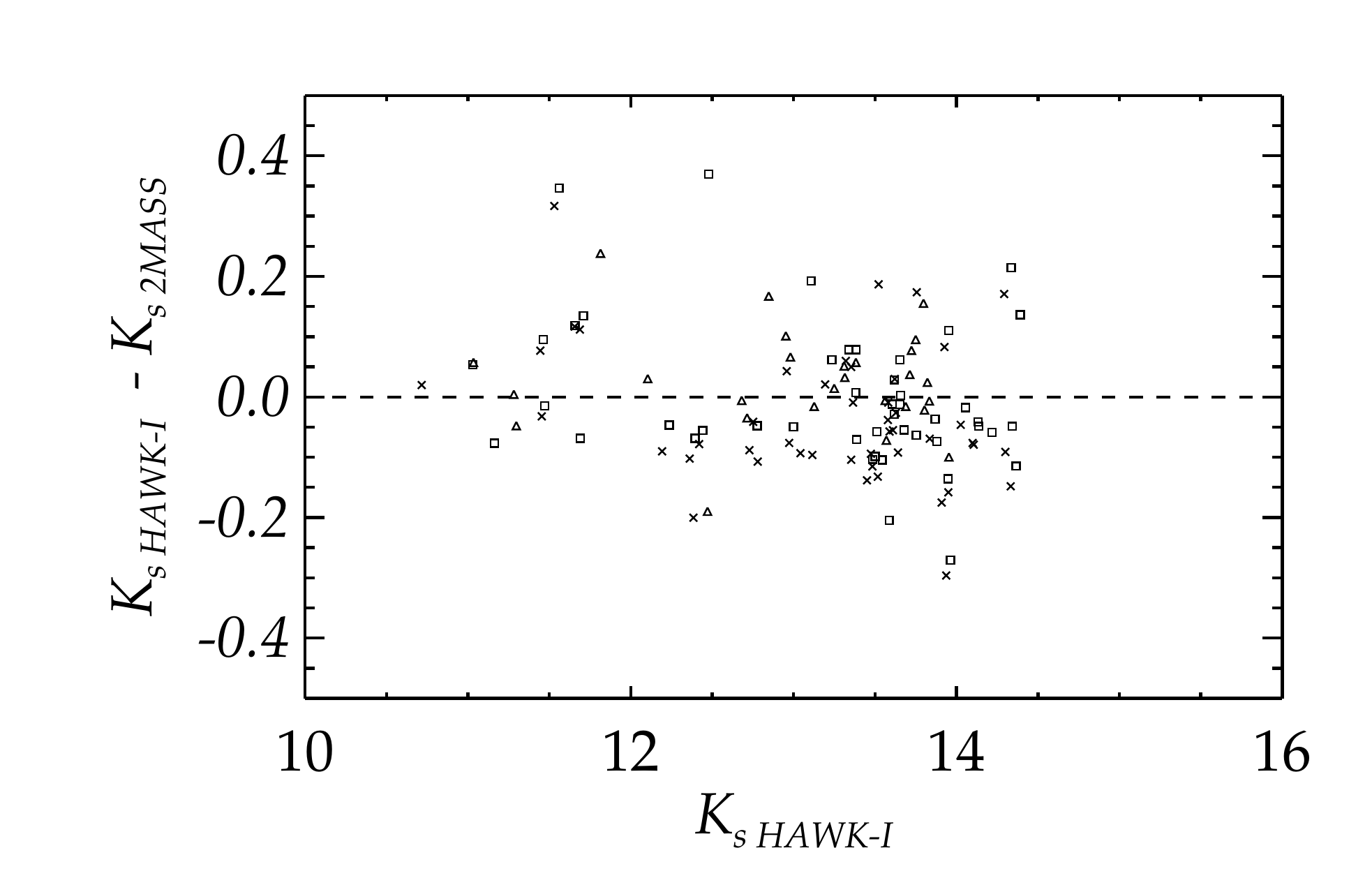}
\caption{Residuals of the calibrated HAWK-I \ks-band photometry compared with the 
2MASS values.  Field~1: open squares; 
Field~2: open triangles; Field~3: crosses.}\label{phot_cf1}
\end{center}
\end{figure}

One \ks-band HAWK-I array with a reasonable spatial overlap was selected for each
MAD image.  These frames were then reduced and
analysed using the same methods as for the MAD images,
i.e. using PSF-fitting photometry.  There were $\sim$50 stars from 2MASS with `AAA'
$J$\h\ks\/ quality ratings in each of the HAWK-I frames.  The
resulting \ks\/ ZPs were applied to all of the objects found in the
image and a catalogue was created.  A comparison of the HAWK-I
magnitudes with those from 2MASS for the calibration stars is shown in
Figure~\ref{phot_cf1}, with dispersions of $\pm$0.12, 0.16, and
0.21$^{\rm m}$ for Fields~1, 2 and 3, respectively.

These results provided a well-defined ZP between HAWK-I and 2MASS,
which was then used to bootstrap the MAD ZPs from stars overlapping in
the images. By comparing the instrumental magnitudes of 30 stars in
the MAD frames with their corresponding HAWK-I values, new ZPs were
calculated for each \ks\/ MAD field. Note that the stars used were
hand-picked to be isolated and well within the dynamic range of the
MAD observations -- using fainter sources from the MAD data would have
pushed toward the sensitivity limits of the HAWK-I frames rather than
providing improved calibration of the MAD data.  The internal
agreement of overlapping stars within the three HAWK-I frames used for
calibration of the three MAD fields was found to be better than
0.1$^{\rm m}$.

Stars that appear in the overlap regions of the MAD fields were used
as a double-check of the ZPs from the HAWK-I data.  Fields~1 and 3
were in excellent agreement, while the ZP for Field~2 gave magnitudes
that were 0.07$^{\rm m}$ brighter, although still within the calculated
uncertainties; the ZP for Field~2 was corrected accordingly, resulting in 
\ks-band magnitudes on a common system.

Unfortunately there were no matching HAWK-I observations in the
\h-band to employ a similar calibration method. Informed by the
behaviour of the \ks-band results, we adopt the 2MASS ZPs from
Table~\ref{zero_points} for Fields~1 and 3, which have much smaller
standard deviations than found for Field~2. Instrumental magnitudes
were then compared for stars overlapping between Fields~1 and 3 as
done for \ks; the resulting offsets were in excellent agreement with
those expected from the ZPs from Table~\ref{zero_points} (i.e.,
$\Delta$\h$\sim$0.16$^{\rm m}$).  We then compared instrumental
magnitudes for stars overlapping between Fields~1 and 2.  These
offsets were then used to calculate a new mean ZP for Field~2, which
is within the quoted uncertainties from the 2MASS comparison, but
has a much reduced error.

\subsubsection {PSF Fitting on Individual Frames}\label{indiv}

If the PSF of a star varies subtly between different frames (as is
likely with AO-corrected imaging) the resulting combined PSF may
broaden slightly and be noisier.  To investigate if this effect was
the dominant source of our ZP uncertainties, we employed a second
photometric method to see if we could improve the calibrations. We
used PSF-fitting subtractions on the individual frames, then used
Dr.~Peter Stetson's {\sc daomatch} and {\sc daomaster} packages to
create a mean catalogue.  This method potentially provides more
precise PSF fitting, leading to improved fidelity of the final
photometric catalogue. It also allows the user to set the number of
frames in which a source must be detected for its inclusion in the
final catalogue.  This has the big advantage that objects in the
dithered regions (previously discarded when the combined image is
subset) can now be included, increasing the spatial extent of the
final source catalogue, and also leads to fewer rejections of stars
affected by cosmics or bad pixels in only one frame.

\begin{table}
\begin{center}
\caption{Photometric zero-points obtained from the individual
frame method. As in Table~\ref{zero_points}, the \ks-band results are derived
from HAWK-I, with the \h-band results from 2MASS.  In both cases Field~2 was
calibrated using stars overlapping with Field~1.}\label{final_zero_points}
\begin{tabular}{lcccc}
\hline
 Band & Field 1 & Field 2 & Field 3 \\
\hline
\ks  &  23.40 $\pm$ 0.07 & 23.62 $\pm$ 0.08 & 23.59 $\pm$ 0.11 \\
\h   &  24.45 $\pm$ 0.14 & 24.37 $\pm$ 0.14 & 24.33 $\pm$ 0.08 \\
\hline
\end{tabular}
\end{center}
\end{table}

The uncertainties on the ZPs obtained from analysis of the individual
frames were (effectively) unchanged compared to those found from our previous best
efforts (see Table~\ref{final_zero_points}
cf. Table~\ref{zero_points}).  This suggests that the uncertainties in
the magnitudes of the calibration stars are dominanting those from the
photmetric methods employed.  A more tangible gain is the increased
number of sources detected due to the retention of the dithered
regions not observed in all frames (albeit with lower sensitivity).
Following application of the ZP calibrations, a comparison between the
combined-frame and individual catalogues was undertaken for each pointing/band, 
finding mean offsets less than 0.05$^{\rm m}$ in all six instances.

\begin{figure}
\begin{center}
\hspace{-0.85cm}\includegraphics[height=6cm]{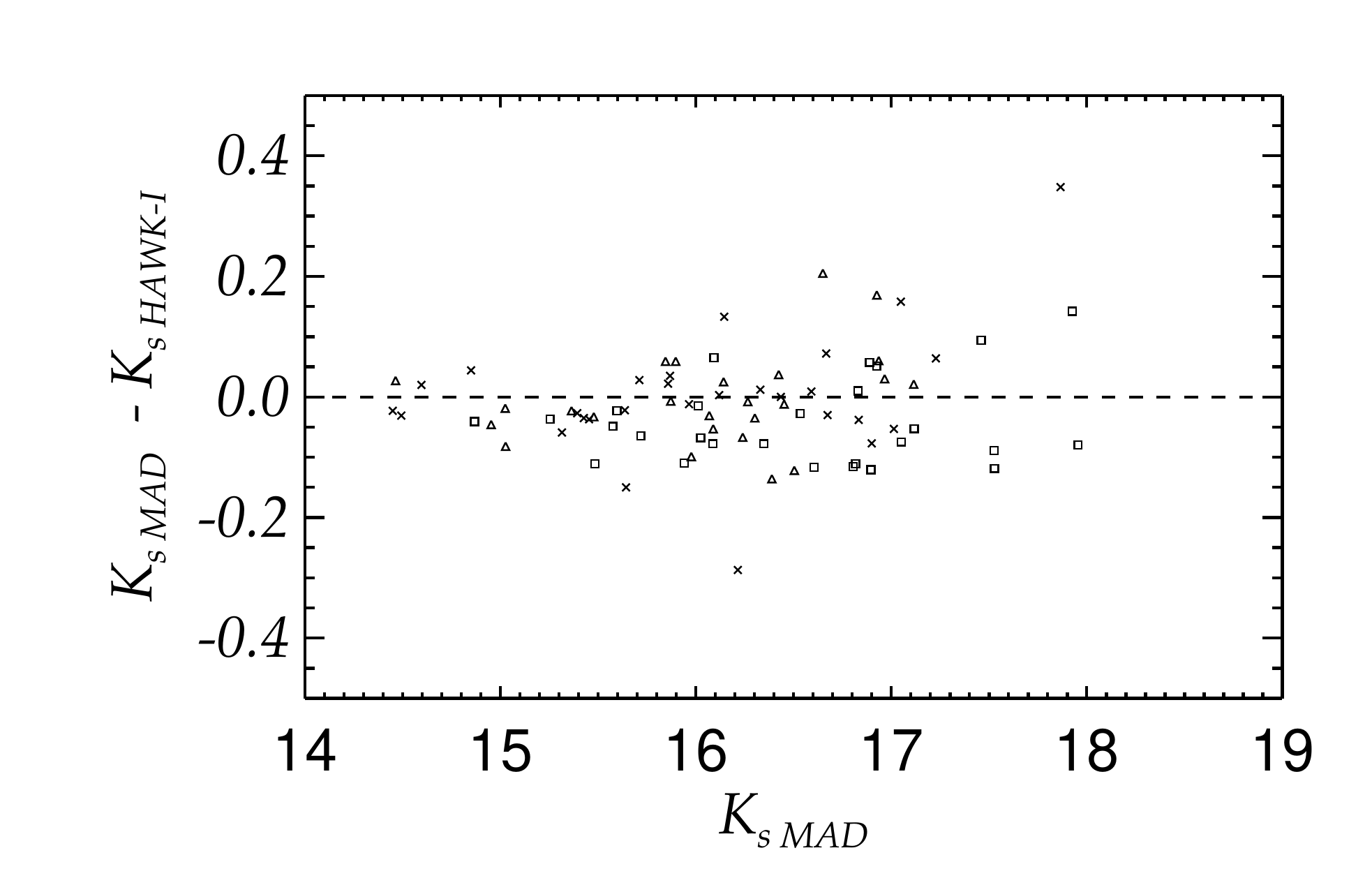}
\caption{Residuals of the calibrated MAD \ks-band photometry from the individual frame
method compared with the HAWK-I values for hand-picked reference
stars (see text for details)}.\label{phot_cf2}
\end{center}
\end{figure}

The ZPs adopted for calibration of the MAD fields for the
final catalogue, created from the individual frames, are listed in
Table~\ref{final_zero_points}. In summary, the \ks-band ZPs were calibrated using
HAWK-I, with Field~2 bootstrapped using overlapping stars in
Field~1. The \h-band ZPs were calibrated using 2MASS, also with
Field~2 adjusted from comparisons with Field~1.
Figure~\ref{phot_cf2} compares the final MAD magnitudes with
those of the HAWK-I calibration stars, with the dispersions around the
mean quoted in Table~\ref{final_zero_points}.

As an external check on the MAD magnitudes we compared our
results for the 12 (visually-matched) overlapping stars in the
NICMOS observations from \citet{br01}.  Nine of these stars were in Field~2 (with two
only observed in the \h-band due to the dither pattern), with the
remaning three in Field~3.  Accounting for the slight differences in
the photometric filters (the {\em HST} data were transformed to the
CTIO photometric system), the mean differences (MAD\,$-$\,{\em HST})
and standard errors are in excellent agreement
$\Delta$\h\,$=$\,$-$0.05\,$\pm$\,0.03 and
$\Delta$\ks\,$=$\,$-$0.04\,$\pm$\,0.05 (std. err.).

\subsection {Colour-magnitude diagrams}

Colour-magnitude diagrams (CMDs) were created from catalogues from
both photometric methods (i.e. from the combined and the individual
frames), excluding objects within a radius of 2\farcs8 of the centre of R136,
where crowding/blending become dominant.

The individual frame method yields an extra 600-1,000 objects in the
CMDs compared to the analysis of the combined frames, mainly due to
the increase in the effective field-of-view.  When CMDs from the
individual method are compared over the same spatial region as the
combined CMD, Fields~2 and 3 contain slightly fewer objects,
predominantly at fainter magnitudes, suggesting the individual
frame method results in slightly diminished sensitivity.  In contrast,
Field~1 still contains fewer objects in the combined CMD, which is a
legacy of a light leak down one side of the images that affected some
of the MAD SD observations.  Field~1 was the
most affected by this light leak in our observations, with the image quality parameters in
{\sc daophot} rejecting more sources in analysis of the combined frame
in this region than in the individual frames.

\begin{figure}
\begin{center}
\hspace{-0.8cm}\includegraphics[width=8.5cm]{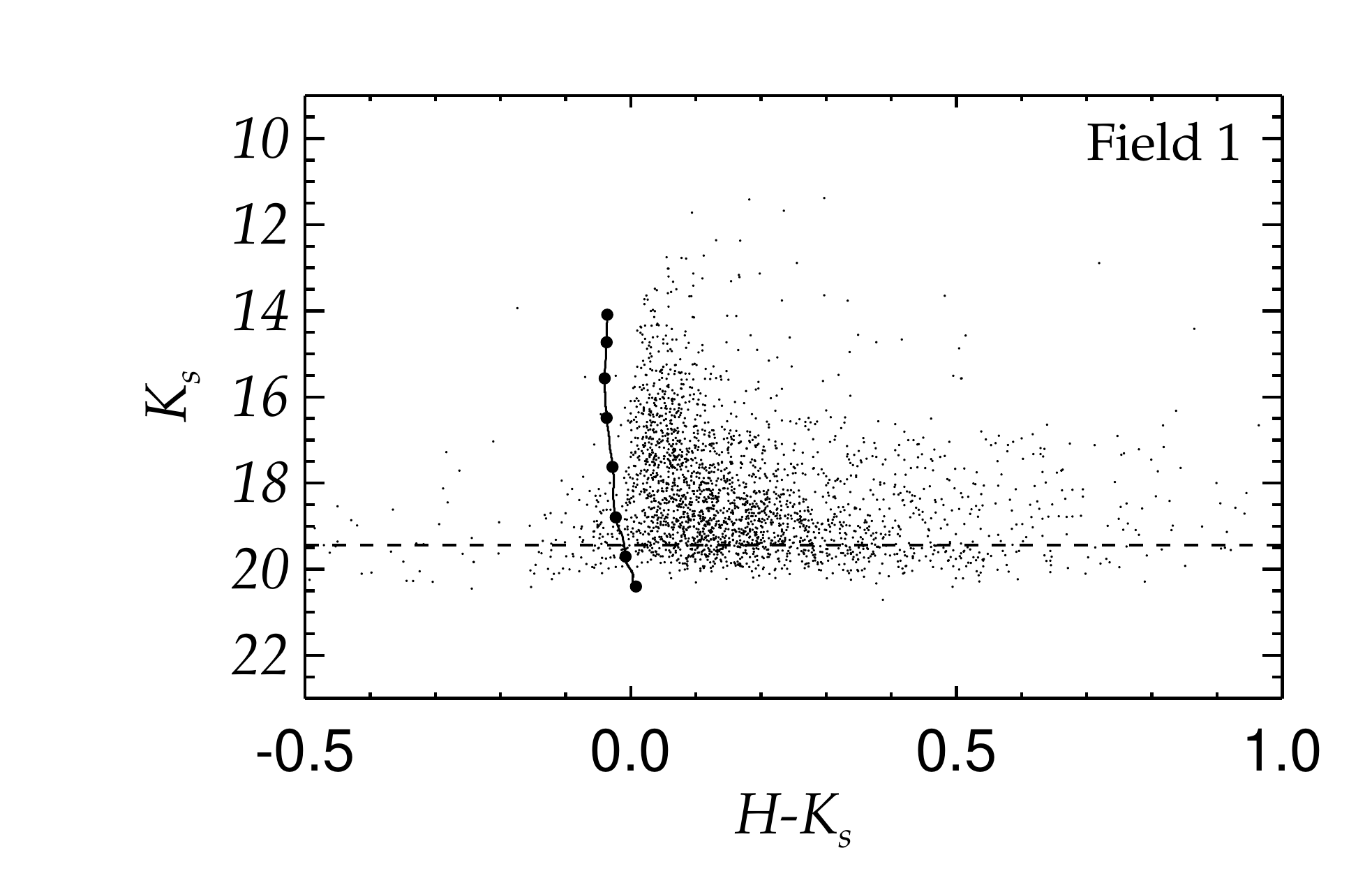}\\
\vspace{-0.35cm}\hspace{-0.8cm}\includegraphics[width=8.5cm]{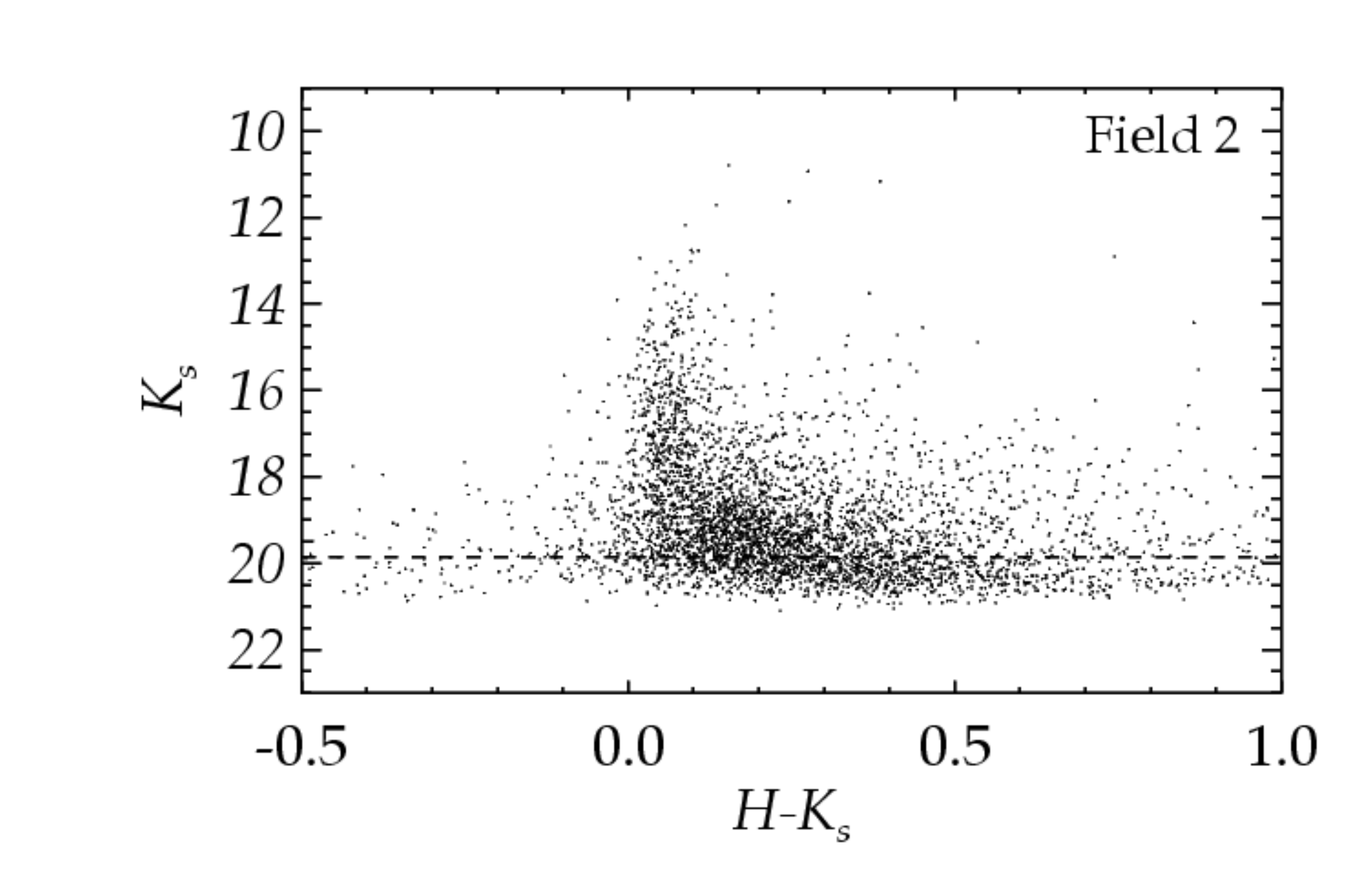}\\
\vspace{-0.35cm}\hspace{-0.8cm}\includegraphics[width=8.5cm]{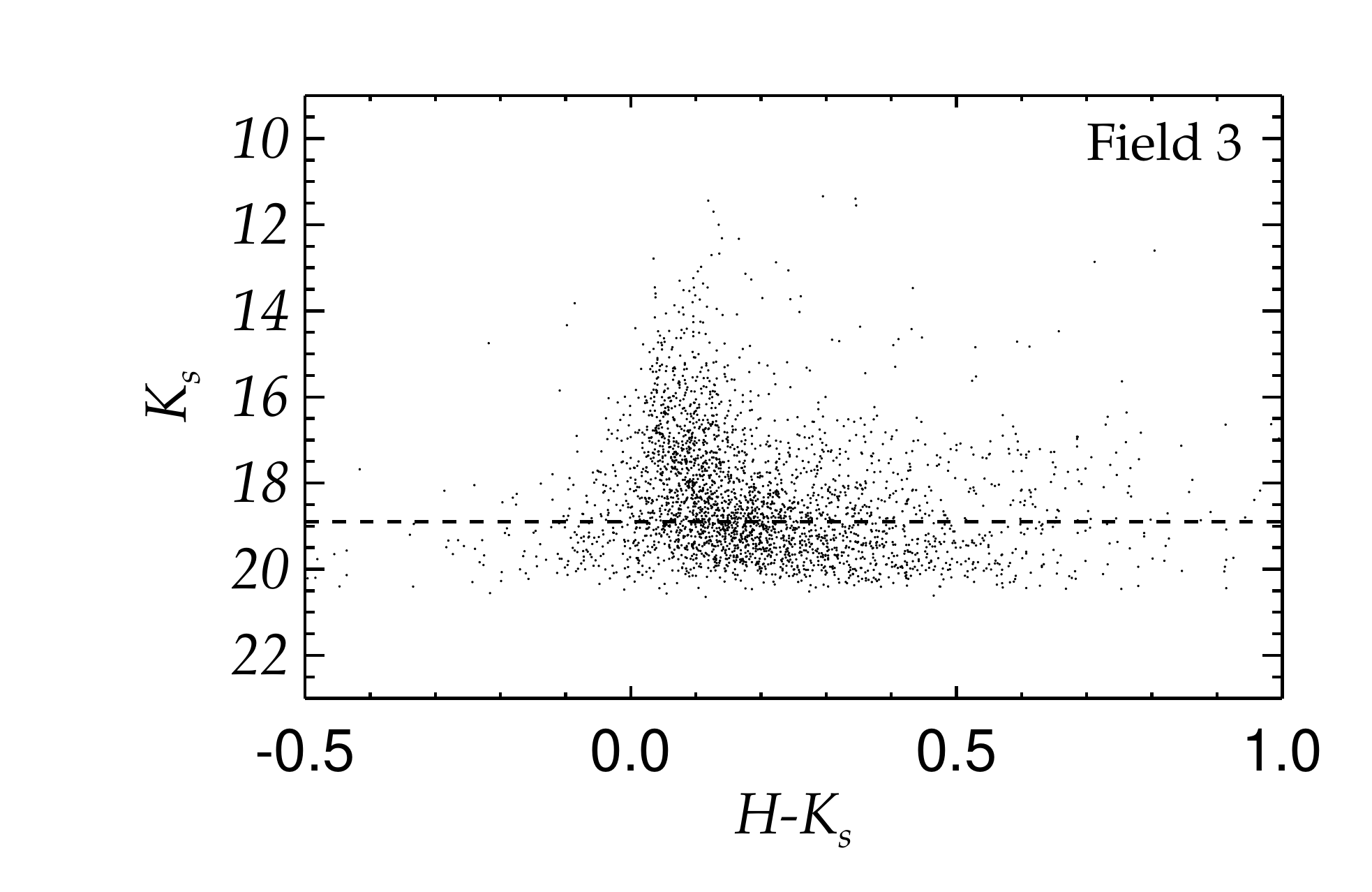}
\caption{CMDs from the mean catalogues created from the individual frames method, 
excluding the central 2\farcs8 of R136.  The dashed horizontal
lines mark the 50\% completeness level (cf. Figure~\ref{complete}) 
in the common regions included in all positional dithers. Overplotted in the
Field~1 panel is the {\em unreddened} youngest isochrone 
(log$t$\,$=$\,3.0) from \citet{ls01}, with the marked points indicating, 
from bright to faint magnitudes, initial masses of 60, 50, 25, 15, 9, 5, 3, and 2\,M$_\odot$.}\label{CMDs}
\end{center}
\end{figure}

In addition to the inclusion of the dithered regions, the individual
frame method reduces the random star-to-star uncertainties in the
final catalogues.  CMDs from the individual frame method are shown for
all three fields in Figure~\ref{CMDs}. The location of the
main-sequence is in good agreement for all three fields.

There is significant and variable extinction toward R136, with 
\cite{a09} adopting a median extinction
of $A_V$\,$=$\,1.85 for massive stars (M\,$=$\,7-20\,M$_\odot$), 
corresponding to $A_K$\,$\sim$\,0.21
\citep[assuming, for the purposes of an order of magnitude estimate of the
IR extinction, the standard galactic extinction law from][]{rl85} and
$A_H$\,$\sim$\,0.32 \cite[using the scaling from][]{i05}, yielding a
relatively small near-IR extinction term of E(\h$-$\ks)\,$=$\,0.11.
Thus, the \h$-$\ks\/ colours from MAD will, in general, not be
differentially reddened by much more than the photometric errors.
CMDs were created for different radial bins using the combined
catalogue from all three fields and these show no significant
offset in the locus of the main sequence. 

For Field~1 in Figure~\ref{CMDs} we include the unreddened
log$t$\,$=$\,3.0 isochrone (effectively illustrative of the zero-age
main sequence) from \citet{ls01}, adopting the tracks with the
metallicity relevant to that of the LMC \citep[i.e. those
from][]{s93}.  Note that the offset from the MAD data is in agreement
with the reddening estimated above.

\subsection{Photometric Completeness}

Completeness tests were undertaken in both bands for all three
fields. Using the {\sc starlist} and {\sc addstar} routines in {\sc
iraf}, 100 artificial stars with magnitudes varying from 14 to 24 were
distributed uniformly across each combined (and subset) image.  As
before, PSF-fitting was performed, using the same settings and a {\sc
penny}{\small 2} model PSF which is allowed to vary across the field.
Using the known positions of the added stars, the number
recovered from the images within a two-pixel search radius was found.
If more than one object was found within that small radius, the object
with the smallest magnitude difference was considered the artificial
star.  Stars with magnitude differences of greater than $\pm$0.5$^{\rm
m}$ were cut from the detected list, with the intention of excluding
mistaken matches, or artificial stars which have been superimposed on
genuine stars within the image (thus increasing its magnitude).

\begin{figure}
\begin{center}
\hspace{-0.85cm}\includegraphics[height=6cm]{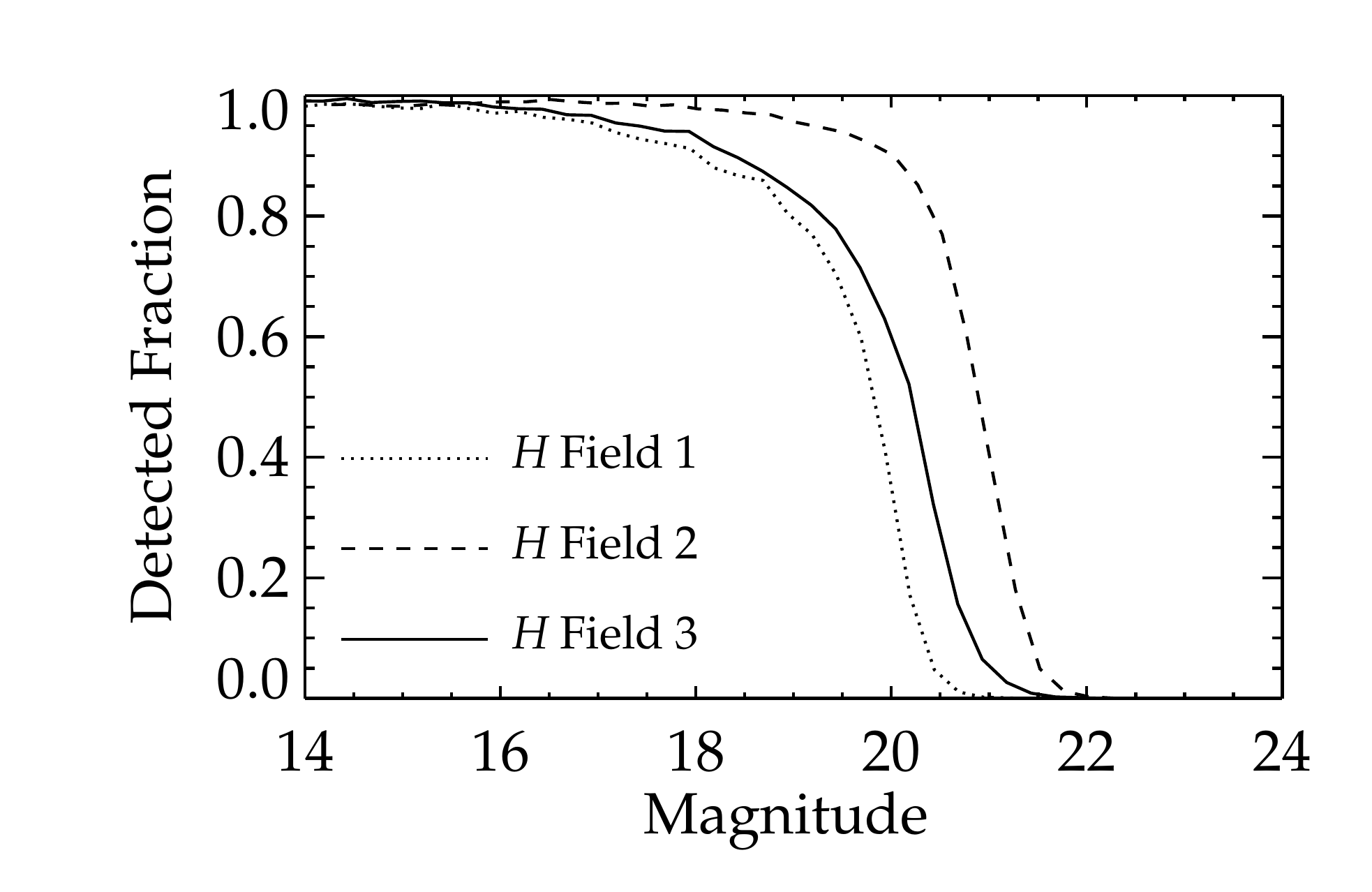}

\hspace{-0.85cm}\includegraphics[height=6cm]{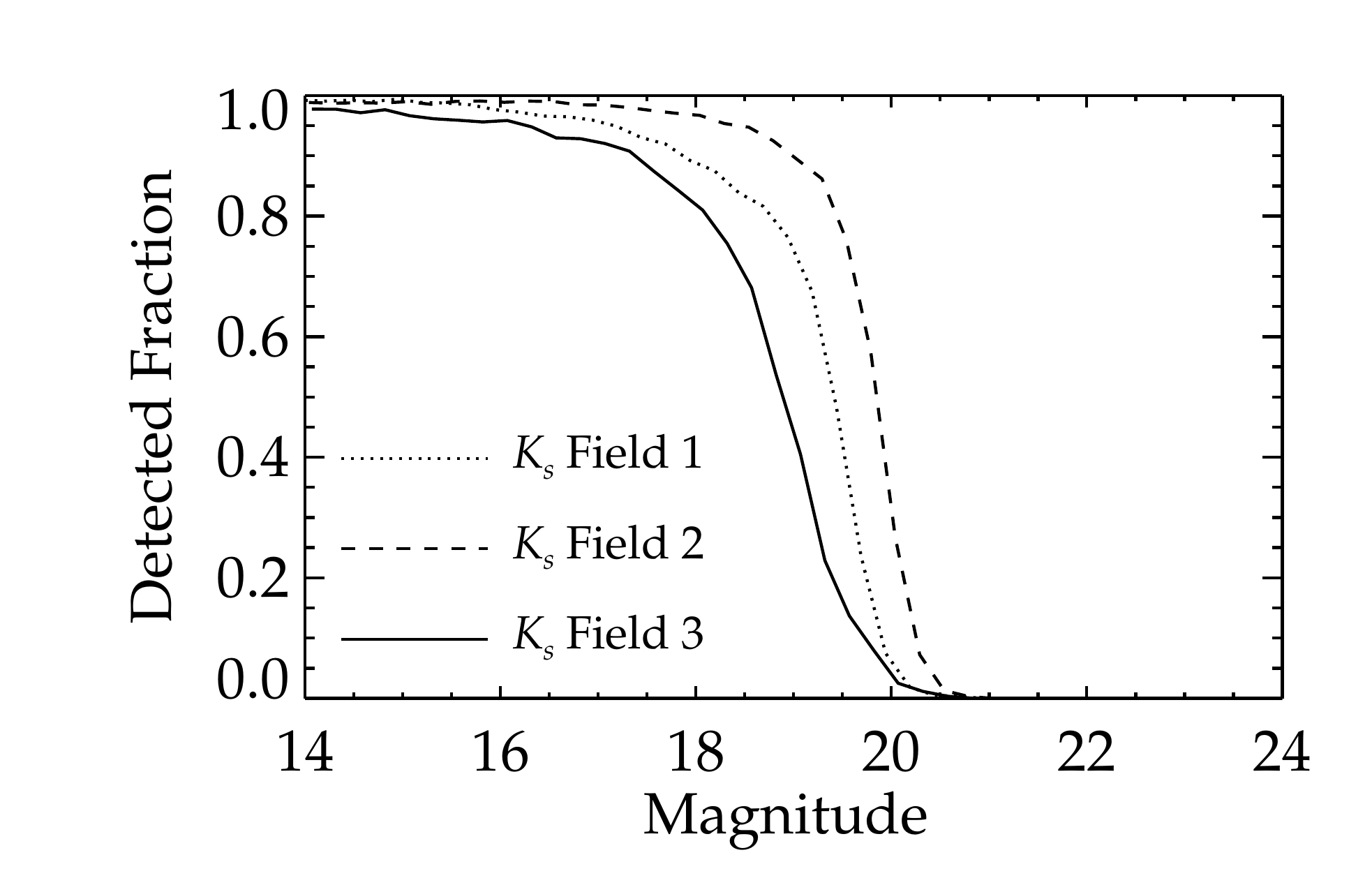}
\caption{\h\/ ({\it upper panel}) and \ks ({\it lower}) completeness profiles for Fields~1, 2 and 3, as 
indicated.}\label{complete}
\end{center}
\end{figure}

These tests were repeated 1,000 times, until 100,000 artificial stars
had been added to each image.  This gives a ratio of $\sim$30 to 50
artificial stars for each observed object.  The ratio of artificial
stars introduced to those detected was measured for each image, as
shown in Figure~\ref{complete}.  Field 2, which does not
include R136 in the combined image, has a slightly more uniform
completeness at brighter magnitudes while going slightly deeper in both bands.

The 50\% completeness level in the \ks-band frames is \ks\,$=$\,19.45 in Field~1, 
19.85 in Field~2, and 18.90 in Field~3, corresponding to initial main-sequence
masses of approximately 3.5, 2.8, and 4.7~M$_\odot$, respectively, when
compared to the youngest isochrone from \citet{ls01}.

\begin{figure*}
\begin{center}
\includegraphics[scale=0.55]{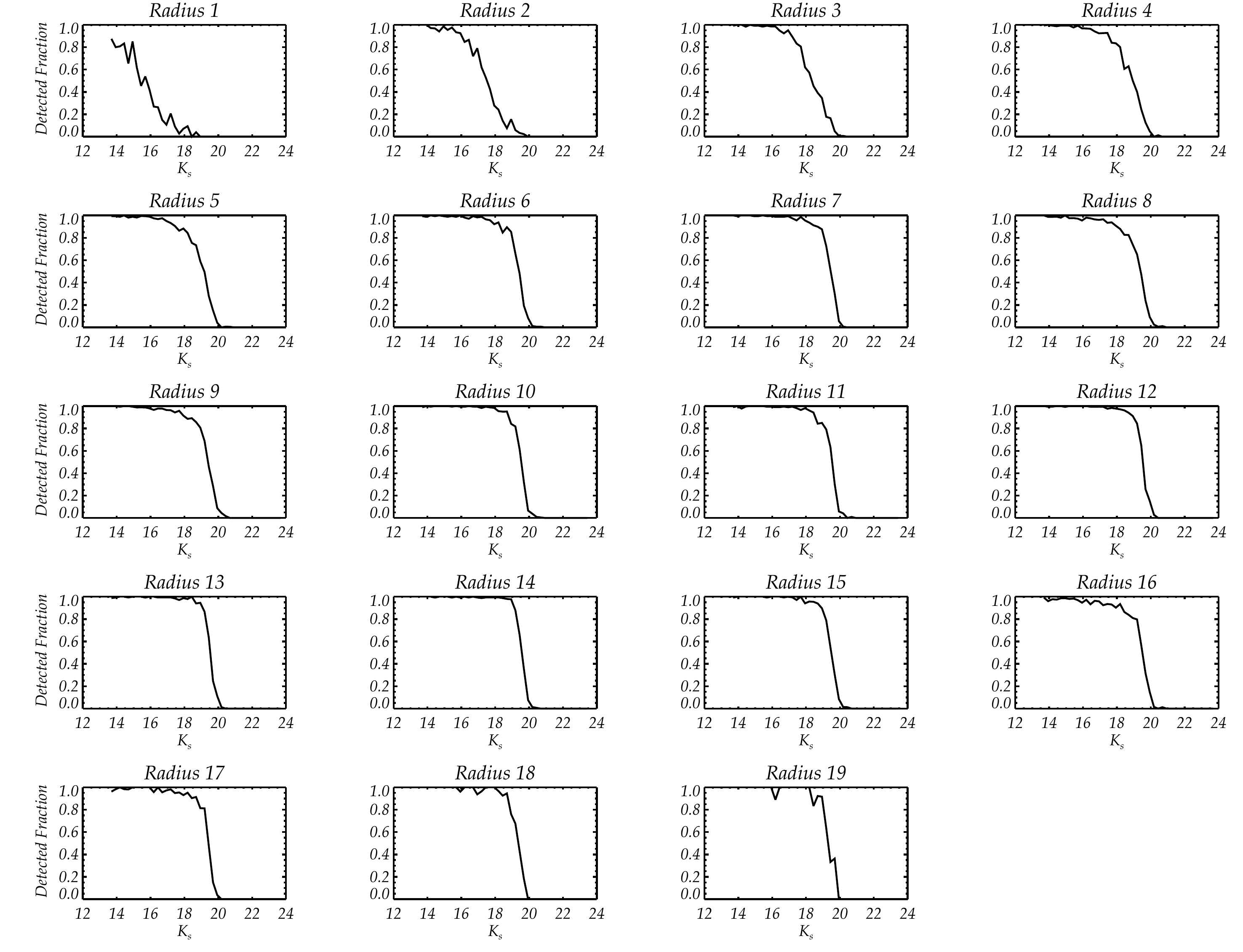}
\caption{\ks-band radial completeness for Field 1. Each annulus has a width of 100 pixels (2\farcs8). 
`Radius~1' refers to the radius closest to the cluster core.}\label{rad_comp_Kp1}
\end{center}
\end{figure*}

For Fields~1 and 3 the completeness was also found for both bands as a
function of radius from the cluster (adopting 100 pixel radial bins),
an example of which is shown in Figure~\ref{rad_comp_Kp1}. These
radial completeness profiles are taken into account in construction
of the radial luminosity profile in Section~\ref{profiles}.

\section{{\it Spitzer} YSO candidates}\label{ysos}
From observations with the {\em Spitzer Space Telescope}, GC09
presented a catalogue of potential young stellar objects (YSOs) in the
LMC.  Prior to investigation of the luminosity profile of R136, we
first investigated the seven candidate YSOs from their catalogue that
are within the MAD fields.  The {\em Spitzer} data were primarily
obtained under the auspices of the Surveying the Agents of a Galaxy's
Evolution (SAGE) LMC survey \citep{mm06}, but were also supplemented
by smaller, more targeted programmes from the {\it Spitzer} archive.

Following initial colour cuts to exclude the majority of stars and
background galaxies, GC09 investigated the nature of the candidate YSOs
via inspection of their morphologies and spectral energy
distributions, including comparisons at other wavelengths.  This
resulted in five categories for the colour-selected {\it Spitzer}
sources: (1) evolved stars, (2) planetary nebulae, (3) background
galaxies, (4) diffuse sources, and (5) `definite', `probable' and
`possible' YSO candidates.  This is a somewhat different approach to
the selection of candidate YSOs from the SAGE LMC survey by
\citet[][see further discussion by GC09]{w08}.  \citet{v09} have since
compared the {\em Spitzer} data with archival {\em HST} H$\alpha$
images and new (seeing-limited) near-IR imaging, enabling more
detailed description of the local environments around 82 of the YSOs
from GC09.

The positional uncertainty on the {\it Spitzer} astrometry is
approximately $\pm$0\farcs2 (Dr~R.~Gruendl, priv. comm.), but in the
crowded environment of 30~Dor, one might expect that blended sources and
nearby gas might lead to greater uncertainties. Indeed, the typical 
angular resolution of the {\it Spitzer} observations ranged
from 1\farcs7 to 2\farcs0 over the four InfraRed Array Camera (IRAC) bands
for the SAGE observations \citep{mm06}.  Thus, it is not surprising that,
in the case of two of the `diffuse' sources from GC09, we find apparent counterparts
in the MAD images offset by a radius of approximately 1$''$.  

In light of these positional uncertainties, there remains one
candidate from GC09 which does not have an obvious close counterpart
within a 2\farcs0 radius: the `diffuse' candidate {\it 053842.69
$-$690623.7}.  There is a faint source (\ks\,$=$\,18.1, \h$-$\ks\,$=$\,0.39) 
at a distance of $\sim$0\farcs75, which is otherwise unremarkable; a brighter
counterpart can be ruled out.  We note that of the four {\em
Spitzer}-IRAC bands, GC09 only reported magnitudes for this source at
4.5 and 8.0\,$\mu$m, suggesting a less robust detection than for the
other six sources discussed here which were detected in all four IRAC
bands.

\begin{figure}
\begin{center}
\includegraphics[width=8cm]{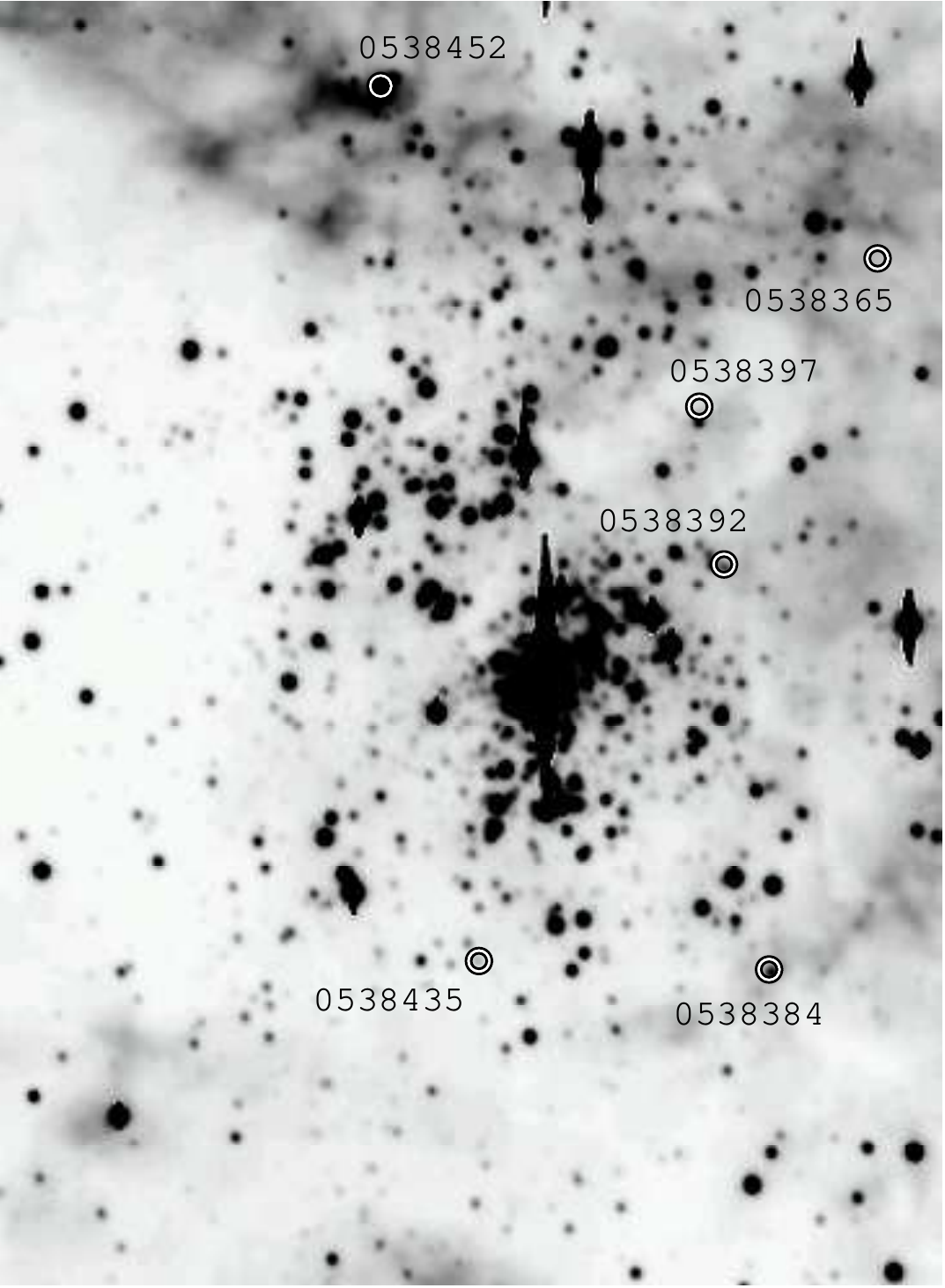}
\caption{Location of the six YSOs from \citet{g09} with counterparts
in the MAD images, overlaid on the $V$-band WFI image.}\label{yso_wfi}
\end{center}
\end{figure}

The remaining six YSO candidates are now discussed in turn, with their
locations overlaid on the $V$-band WFI image in Figure~\ref{yso_wfi}
to illustrate their positions relative to R136.  All except {\it
053836.48 $-$690524.1} feature in the comparison by \citet{v09}, with
four also within the (seeing-limited) near-IR imaging from
\citet[][hereafter RBW98]{rbw98}.  {\em Spitzer} spectroscopy of three
of these sources ({\it 053839.24 $-$690552.3}, {\it 053839.69
$-$690538.1} and {\it 053845.15 $-$690507.9}) was presented by
\citet{s09}, with each classified in their `PE Group', which are seen to display
polycyclic aromatic hydrocarbon (PAH) emission features combined with
fine-structure lines from ionised gas.

\subsection{\it{053836.48 $-$690524.1:}}
\begin{figure}
\begin{center}
\includegraphics[height=3.75cm]{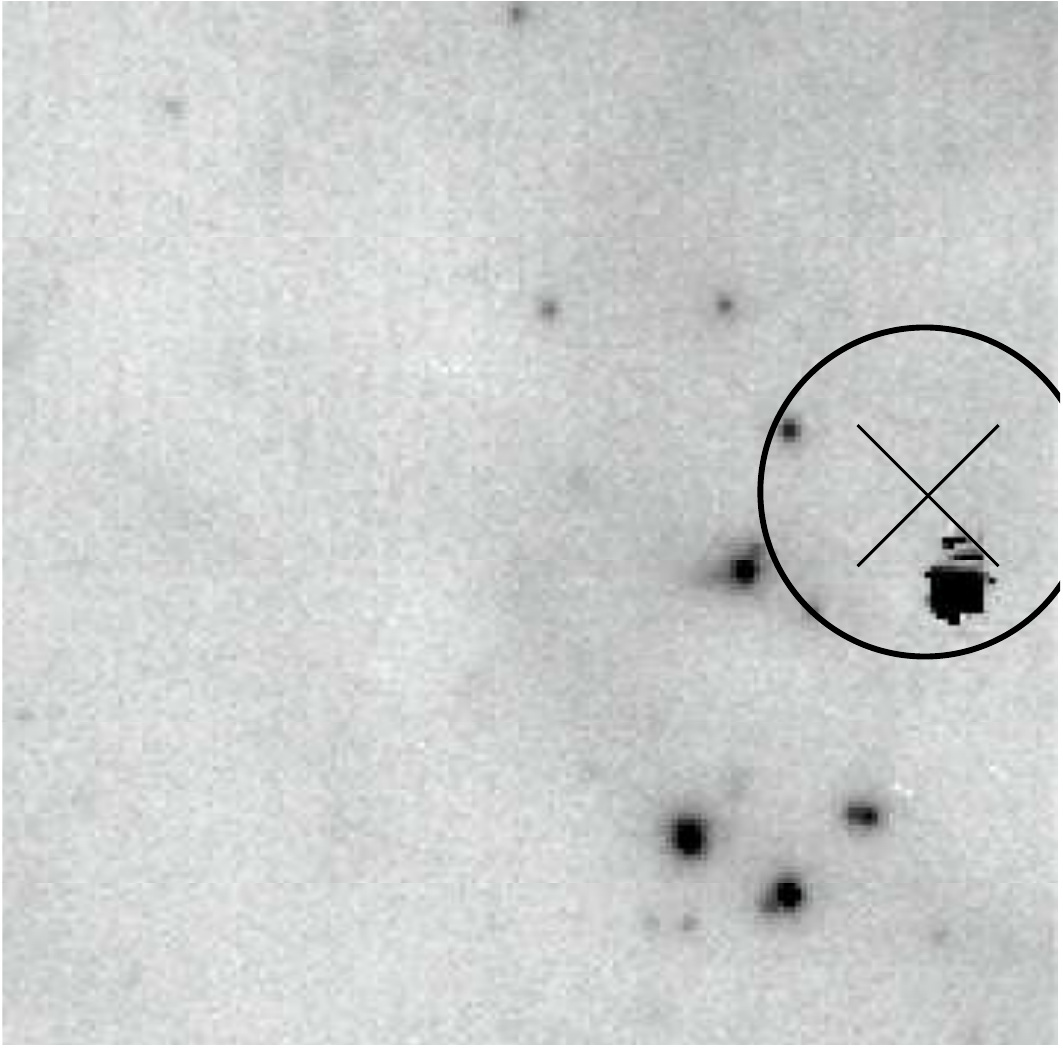}\hspace{0.1cm}\includegraphics[height=3.75cm]{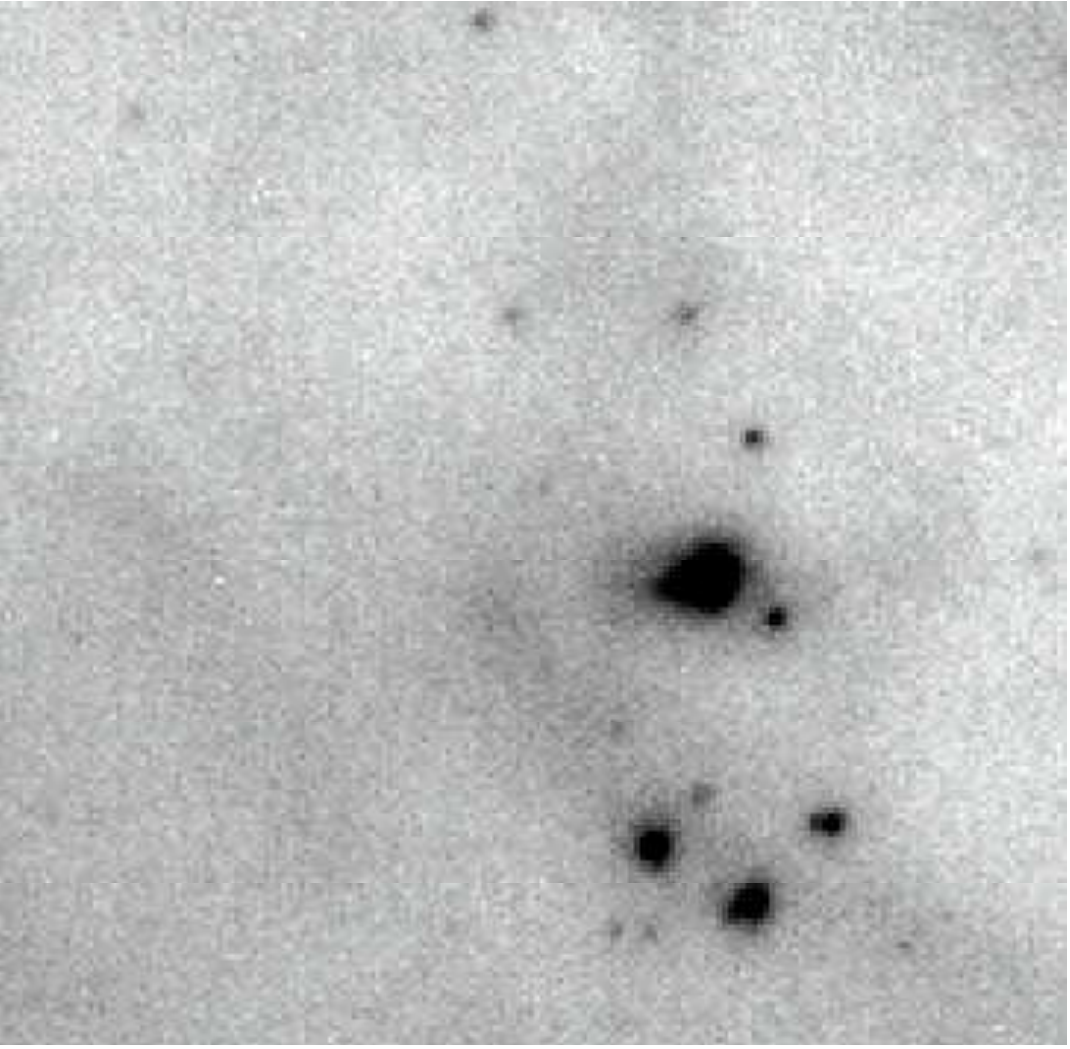}
\caption{053836.48 $-$690524.1: \h- and \ks-band 5$'' \times$5$''$ images
(left- and right-hand panels, respectively), with north at the top,
east at the left.  The black cross marks the location of the {\it
Spitzer} position for the YSO candidate, with the black circle representing
the angular resolution of the IRAC data (1\farcs7 at 3.6\,$\mu$m).}\label{yso1}
\end{center}
\end{figure}

Reported as a `diffuse' source by GC09, this object is on the western
edge of the Field~2 images.  Subset (5$''$\,$\times$\,5$''$) \h\/ and
\ks-band images are shown in Figure~\ref{yso1}, in which the cross in
the \h-band image indicates the position from GC09 and the intensity
scaling is the same in both bands. The circle (centred on the {\it Spitzer}
position in the \h-band image) indicates the typical angular resolution
of IRAC in the SAGE survey, i.e. 1\farcs7 at 3.6\,$\mu$m \citep{mm06}.
Just below the {\it Spitzer} position in the \h-band
image is a small cluster of bad pixels -- the likely YSO counterpart is
to the east and is more clearly seen in the \ks-band image.

Apart from the obvious asymmetry in the \ks-band image, we were
somewhat cautious to match the MAD object to the GC09 source.
However, \citet{kk07} also report a candidate YSO (their
`30\,Dor-15', also from IRAC observations) that is only
$\sim$0\farcs33 from the MAD counterpart, which has \ks\,$=$\,16.24,
\h\,$-$\,\ks\,$=$\,2.08.\footnote{We note the candidate YSO `30\,Dor-22' from
\citet{kk07} is, on the basis of their published astrometry, Mk\,34 \citep{m85}, a
well-studied WN star.  We are puzzled that Kim et al. mention this as
one of the sources studied by \citet{br01}, as it does not feature in their NICMOS fields.}

\subsection{\it{053838.35 $-$690630.4:}}
Classified as a `diffuse source' by GC09, there is a potential counterpart
(\ks\,$=$\,14.37 and \h$-$\ks\,$=$\,0.35) in the MAD images at a distance
of 1$''$ (Figure~\ref{yso2}).  The source appears to have an associated
`plume' of nebulosity which spirals to the south-west for at least
0\farcs5.  This nebulosity can also be seen in the NICMOS image from
\citet[][their 30Dor-NIC01 frame]{br01}, but did not feature in their discussion.

The 2MASS $J$-band magnitude quoted by GC09
for this source is 13.93 (2MASS source: 053838.5-6906297).
Overplotting the 2MASS catalogue on the MAD image reveals reasonable
astrometric agreement (0\farcs2) considering the limited angular
resolution of 2MASS, with upper limits of \h\,$>$\,12.88 and \ks\,$>$\,12.05 (i.e. `U' rated).

This source also corresponds to IRSW-98 from RBW98
\citep[which, in turn, is component `a' of the knot comprising source W4
from][]{rrg92}, who found \ks\,$=$\,14.02 and \h$-$\ks\,$=$\,0.45.
\citet{v09} note that this YSO is in a marginally-resolved H~\2 region, 
suggesting that the line contribution (at least in the \ks-band) is likely
contributing to the brighter magnitude from RBW98 compared to the MAD results.

\subsection{\it{053839.24 $-$690552.3:}}
The most likely YSO counterpart is at $\alpha$\,$=$\,05$^{\rm
h}$\,38$^{\rm m}$\,39\fs28, $\delta$\,$=$\,$-$69$^\circ$\,05$'$\,52\farcs60 (J2000,
$\sim$0\farcs45 from the {\it Spitzer} position) and 
is the most immediately impressive counterpart in the MAD images
(Figure~\ref{yso3})\footnote{The source is on the edge of the 
region common to all \h- and \ks-band images, hence the apparent `banding'
where the sensitivity differs due to the science-frame dithers.}. We find
\ks\,$=$\,14.71 and \h$-$\ks\,$=$\,0.42 for the star, as compared to
\ks\,$=$\,13.91, \h$-$\ks\,$=$\,0.61 from RBW98 (their source IRSW-118).  
The images from RBW98 were taken in typical seeing of 1$''$ so 
the two sources visible a the centre of Figure~\ref{yso3} will be
strongly blended, leading to the brighter magnitude.

Particularly striking is the apparent bow-shock, most easily seen in the 
\ks-band image.  This is almost, but not quite, aligned with the core of
R136 approximately 19\farcs5 away.  Also of note in the same direction
is R134/Brey~75 \citep{b81}, classified as WN6(h) by \citet{cs97}, at
a distance of only 8$''$.  To date, nothing is known regarding the
spectral type of the two bright stars at the centre of the image
(separated by $\sim$0\farcs33).  This region clearly warrants more
detailed investigation in the context of small-scale triggered star
formation.

\subsection{\it{053839.69 $-$690538.1:}}
Classified as a `definite YSO' by GC09, this object appears 
as a somewhat extended or embedded source in the \h-band MAD image 
(Figure~\ref{yso4}), with \ks\,=\,13.75 and \h$-$\ks\,=\,3.37.
This source corresponds to IRSW-127 from RBW98 \citep[W9
from][]{rrg92}, for which they found \ks\,$=$\,13.91 and
\h$-$\ks\,$=$\,2.85.  

The bright star 1\farcs2 to the south is P733/S206, for which
\ks\,=\,14.40 and \h$-$\ks\,$=$\,0.16, suggesting it as a massive star
\citep[cf. the intrinsic colours from][]{mp06}.
The strong contrast in the colours of the two
stars is immediately obvious from the composite-colour image from
RBW98 (their Figure~1), although they appear slightly blended.
Blending could perhaps account for the photometric differences (which are in
the expected direction), but RBW98 employed PSF-fitting methods so
should be relatively robust to such effects at this separation --
perhaps our efforts are somewhat limited by the apparent asymmetric
flux to the south of the star.

\subsection{\it{053843.52 $-$690629.0:}}
Classified as a `possible YSO' by GC09, the nearest spatial match is a source with
\ks\,=\,17.15 and \h$-$\ks\,$=$\,1.01, as indicated in Figure~\ref{yso5}.  The adjacent 
object (0\farcs25 to the northeast) is P1064/S696, with \ks\,$=$\,16.87 and
\h$-$\ks\,$=$\,0.18, and for which the spectral type is unknown.  This YSO source is not 
within the region observed by RBW98 but, interestingly, is noted by \citet{v09} as being 
in a dark cloud and as comprising multiple YSOs in the {\em Spitzer} PSF,
one with an optical counterpart in the H$\alpha$ {\em HST} images, and
one without.

\subsection{\it{053845.15 $-$690507.9:}}
\begin{figure}
\begin{center}
\includegraphics[height=3.75cm]{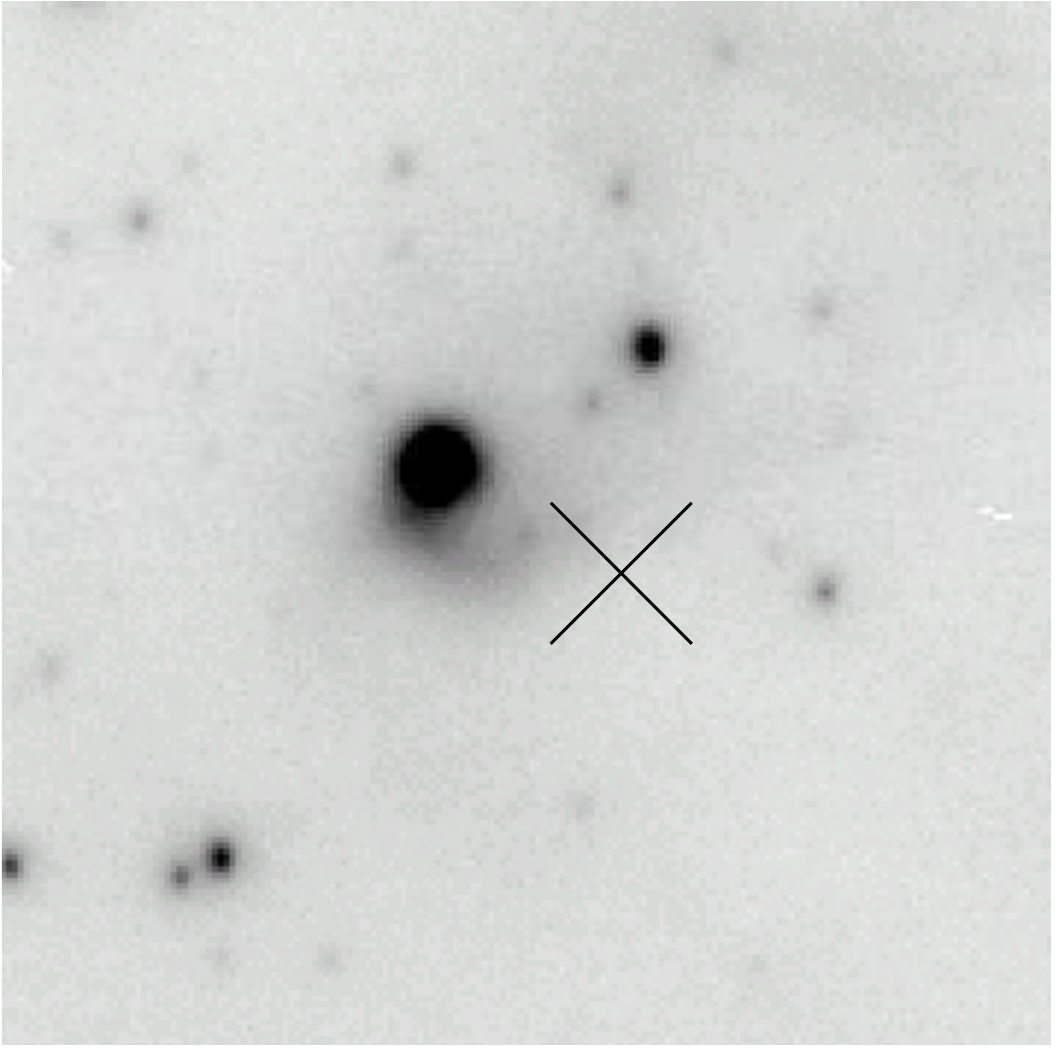}\hspace{0.1cm}\includegraphics[height=3.75cm]{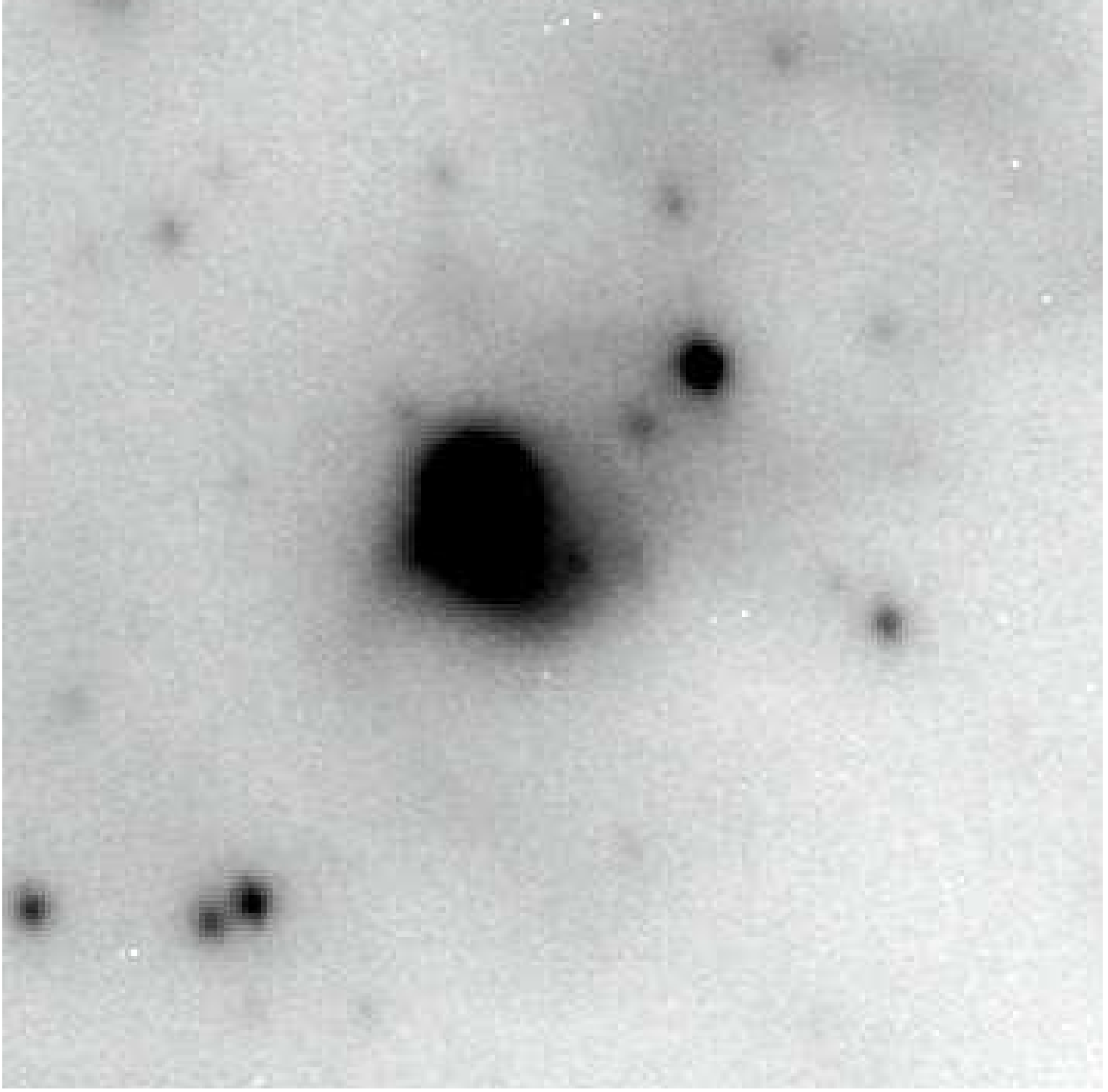}
\caption{053838.35 $-$690630.4: 5$''$\,$\times$\,5$''$ \h- and \ks-band images.}\label{yso2}

\includegraphics[height=3.75cm]{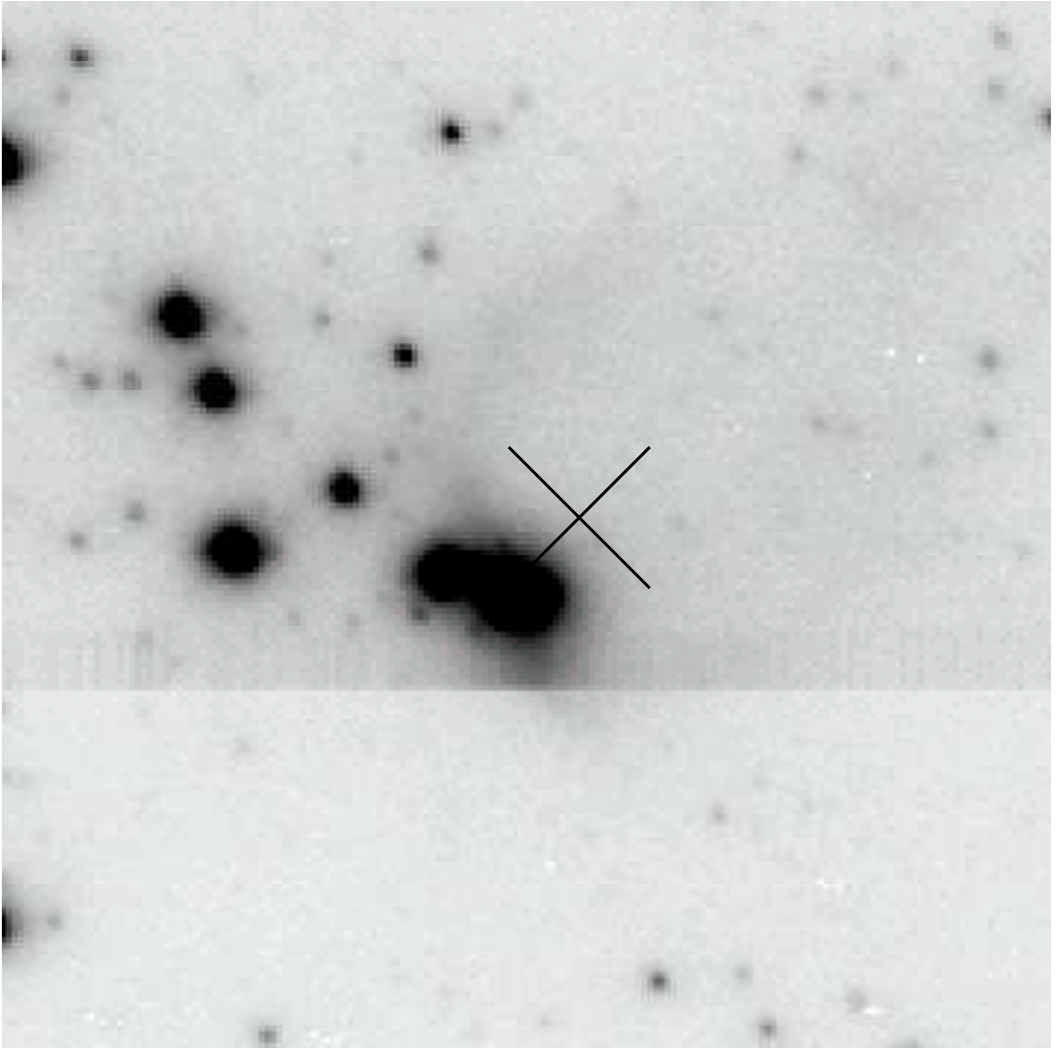}\hspace{0.1cm}\includegraphics[height=3.75cm]{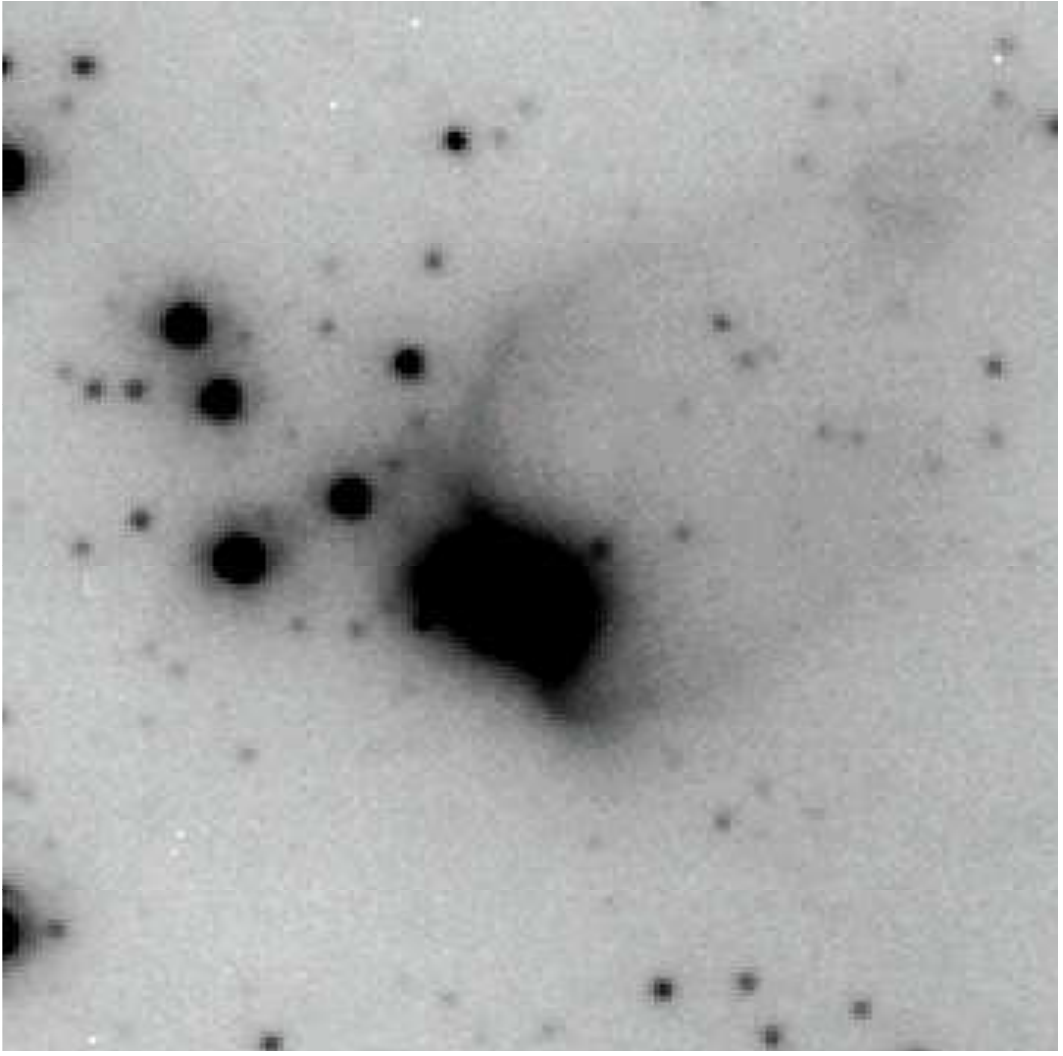}
\caption{053839.24 $-$690552.3: \h- and \ks-band. Note the 
particularly striking bow-shock feature in the \ks\/ image.}\label{yso3} 

\includegraphics[height=3.75cm]{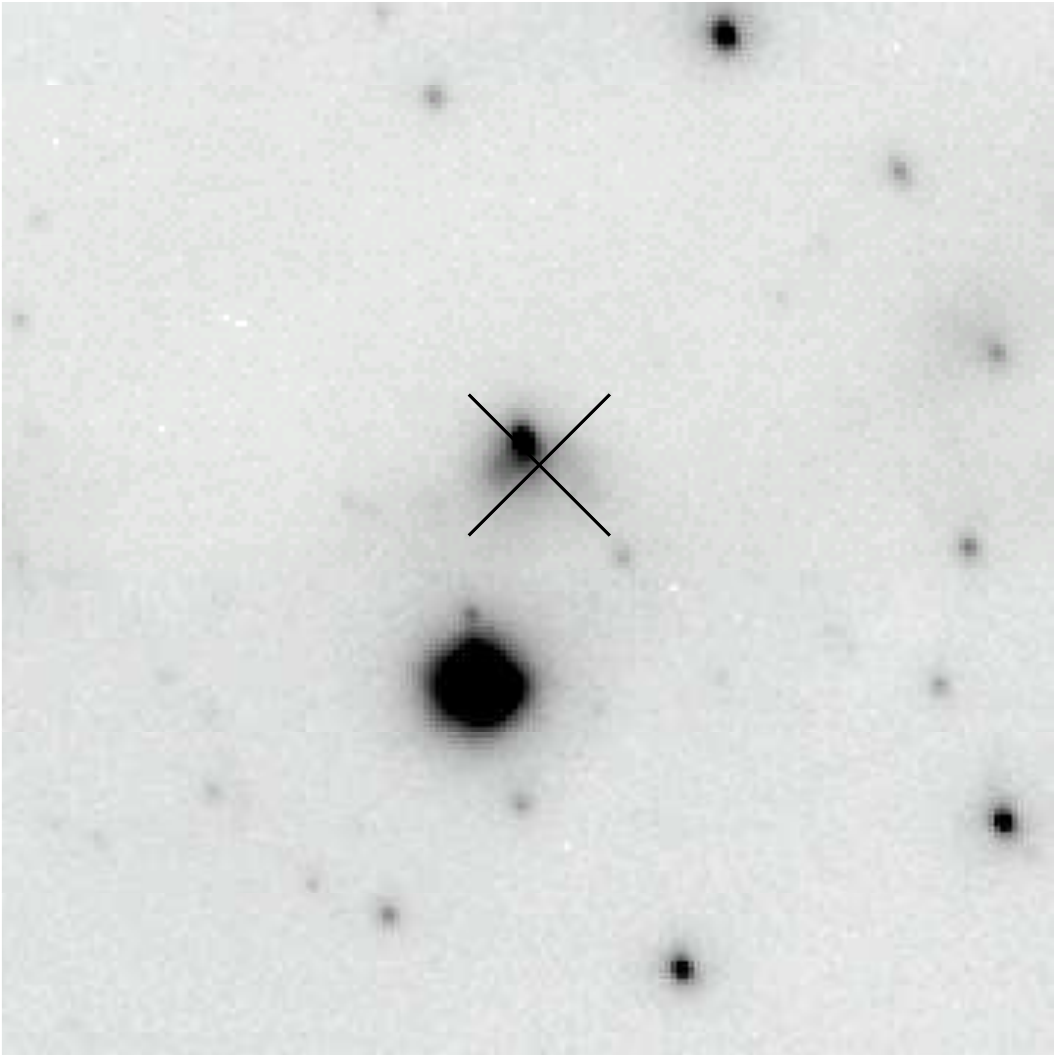}\hspace{0.1cm}\includegraphics[height=3.75cm]{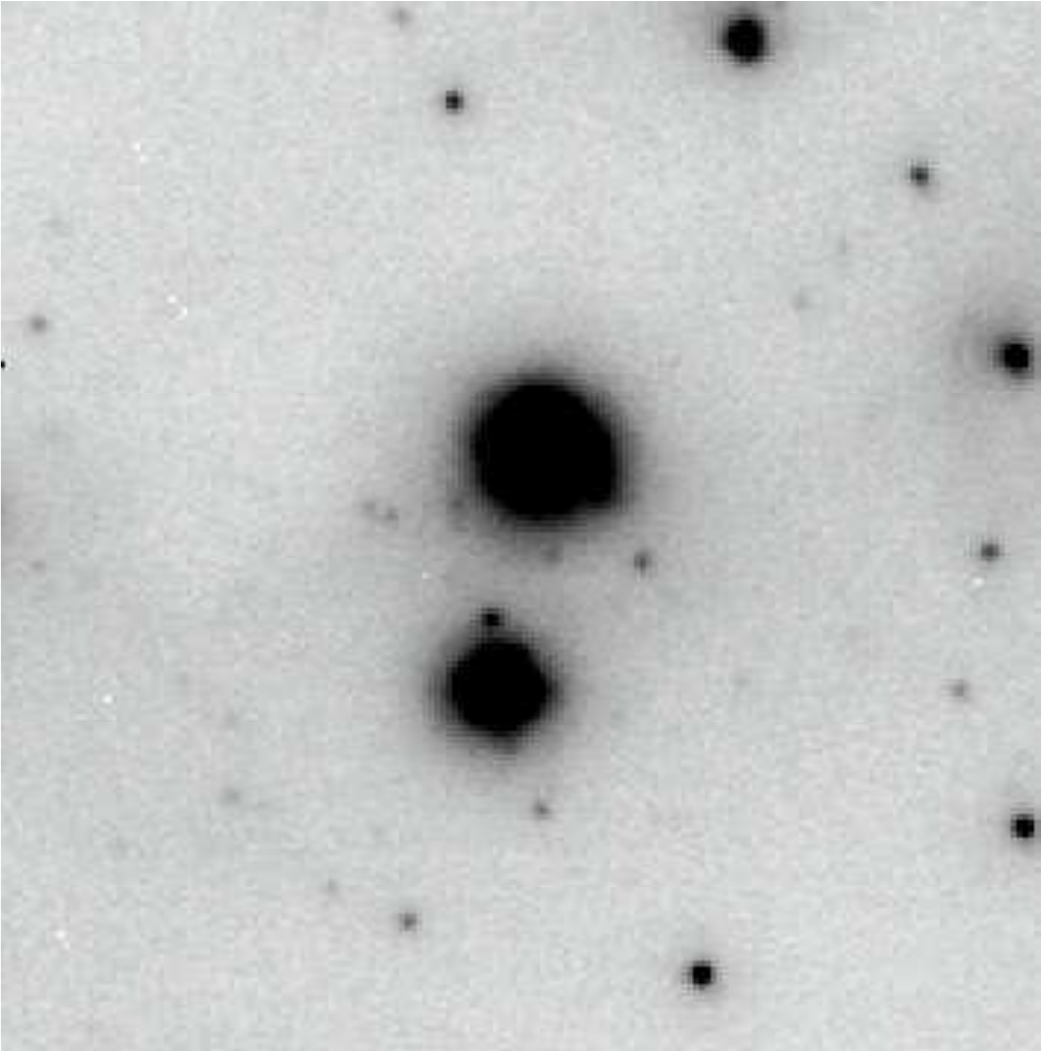}
\caption{053839.69 $-$690538.1: \h- and \ks-band.}\label{yso4}

\includegraphics[height=3.75cm]{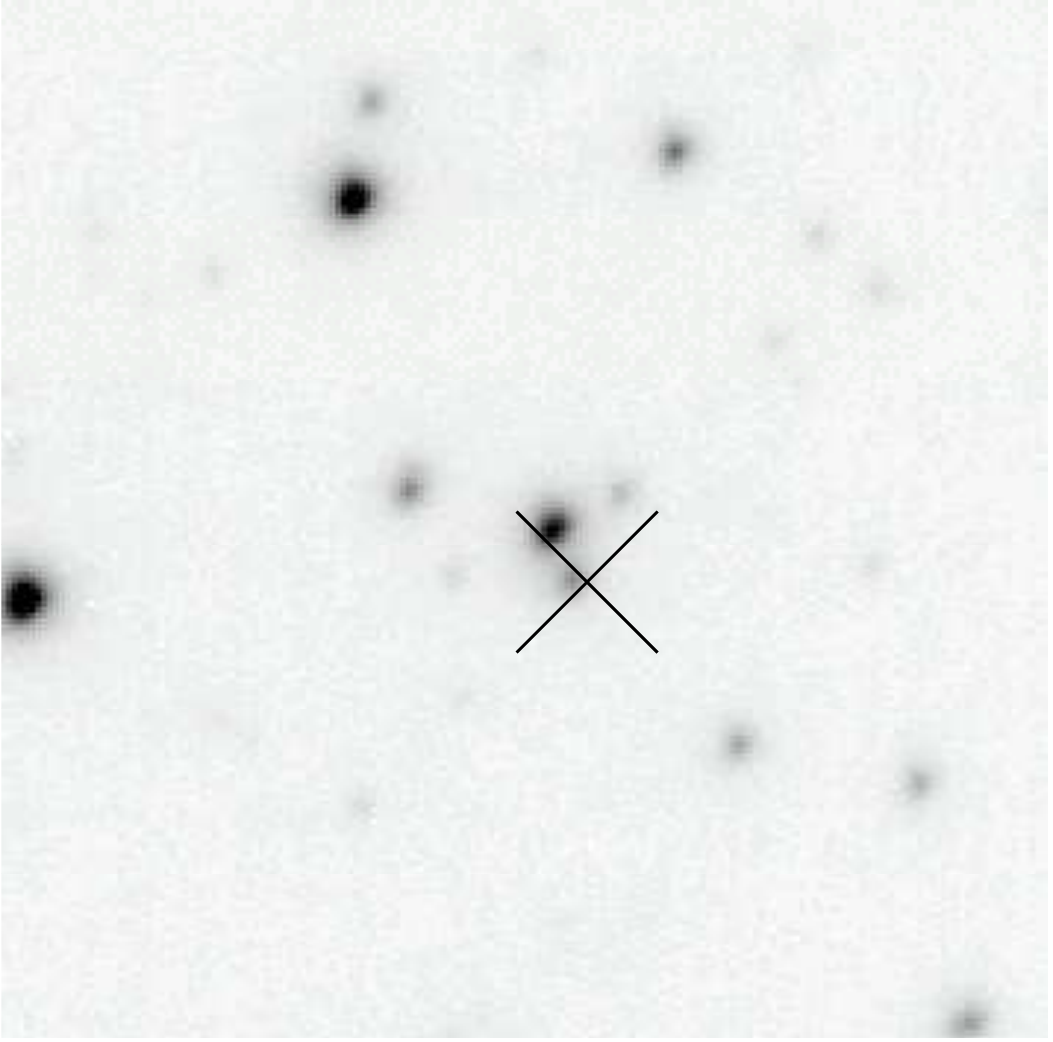}\hspace{0.1cm}\includegraphics[height=3.75cm]{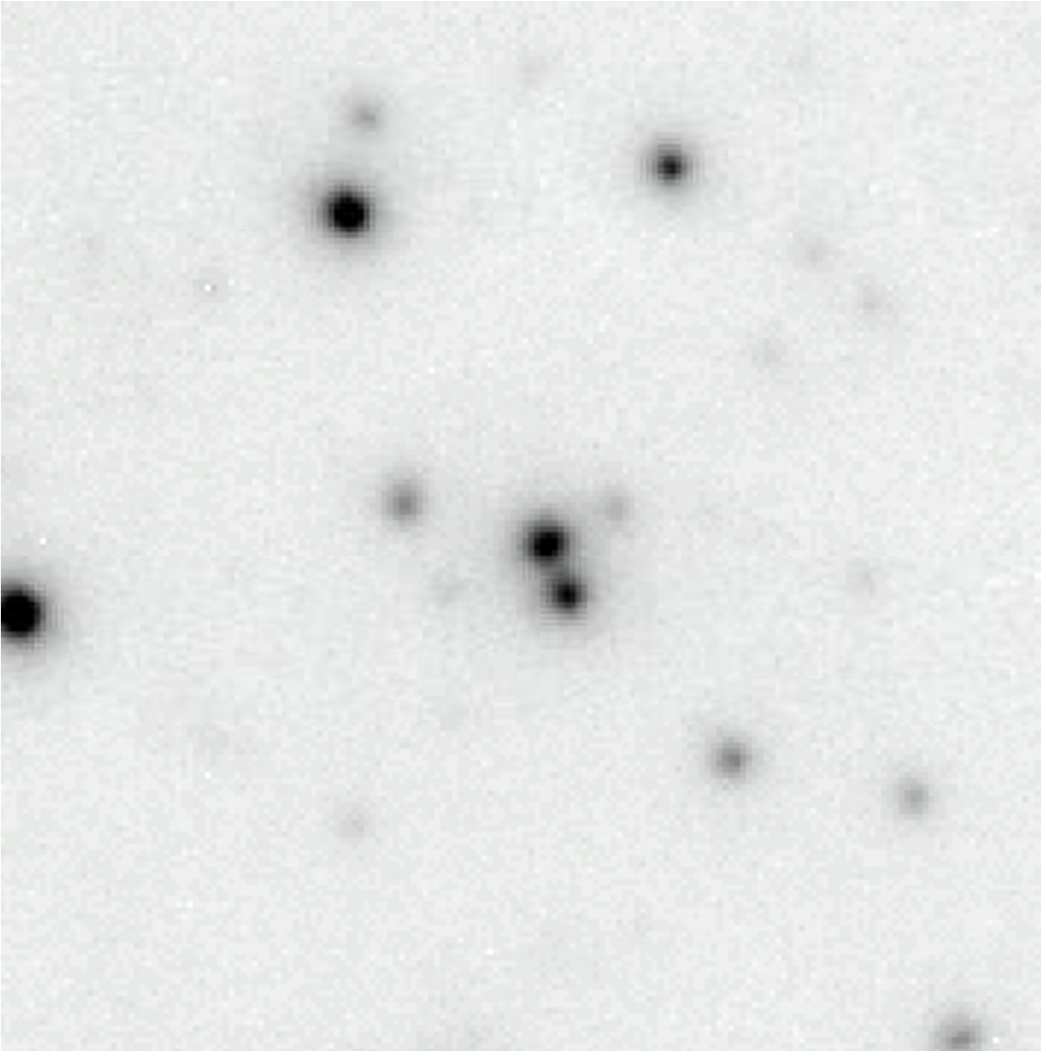}
\caption{053843.52 $-$690629.0: \h- and \ks-band.}\label{yso5}

\includegraphics[height=3.75cm]{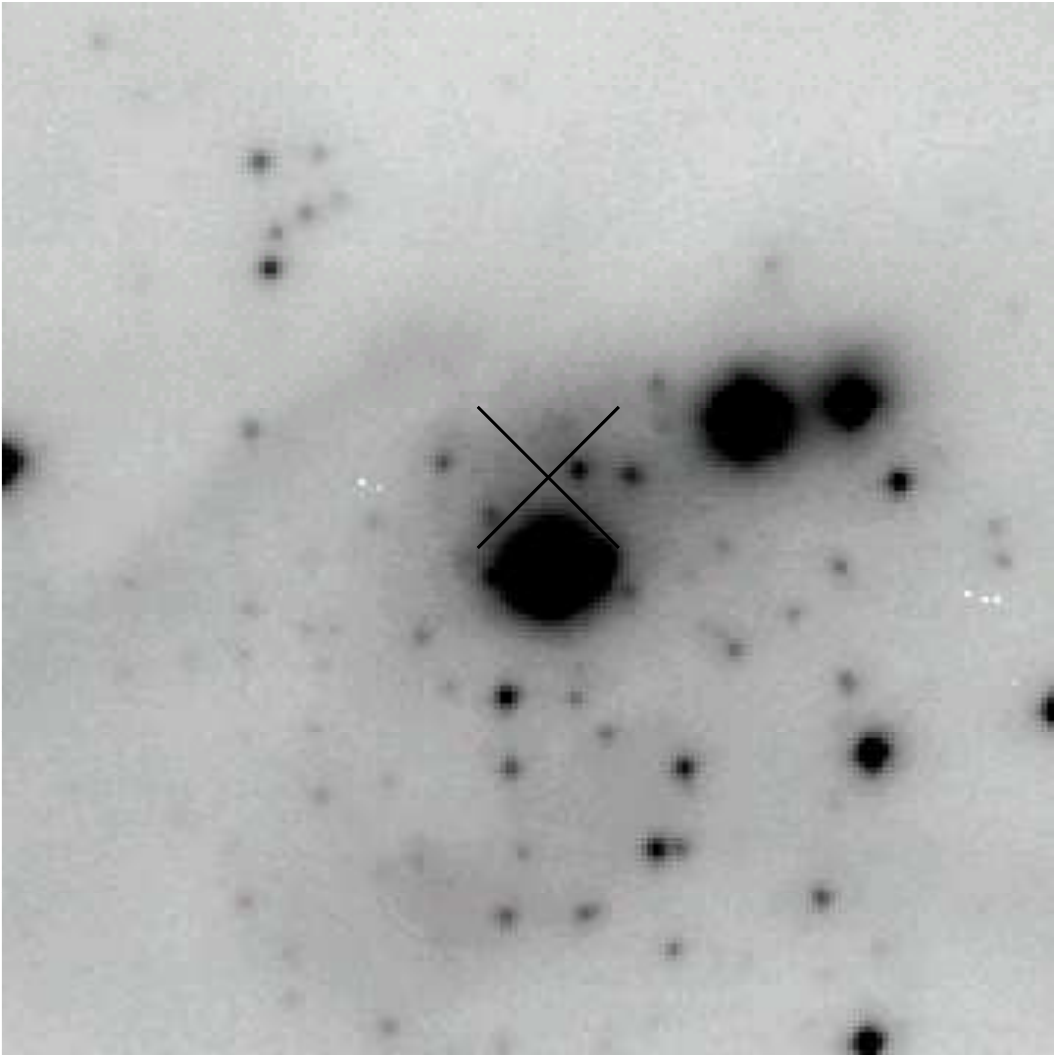}\hspace{0.1cm}\includegraphics[height=3.75cm]{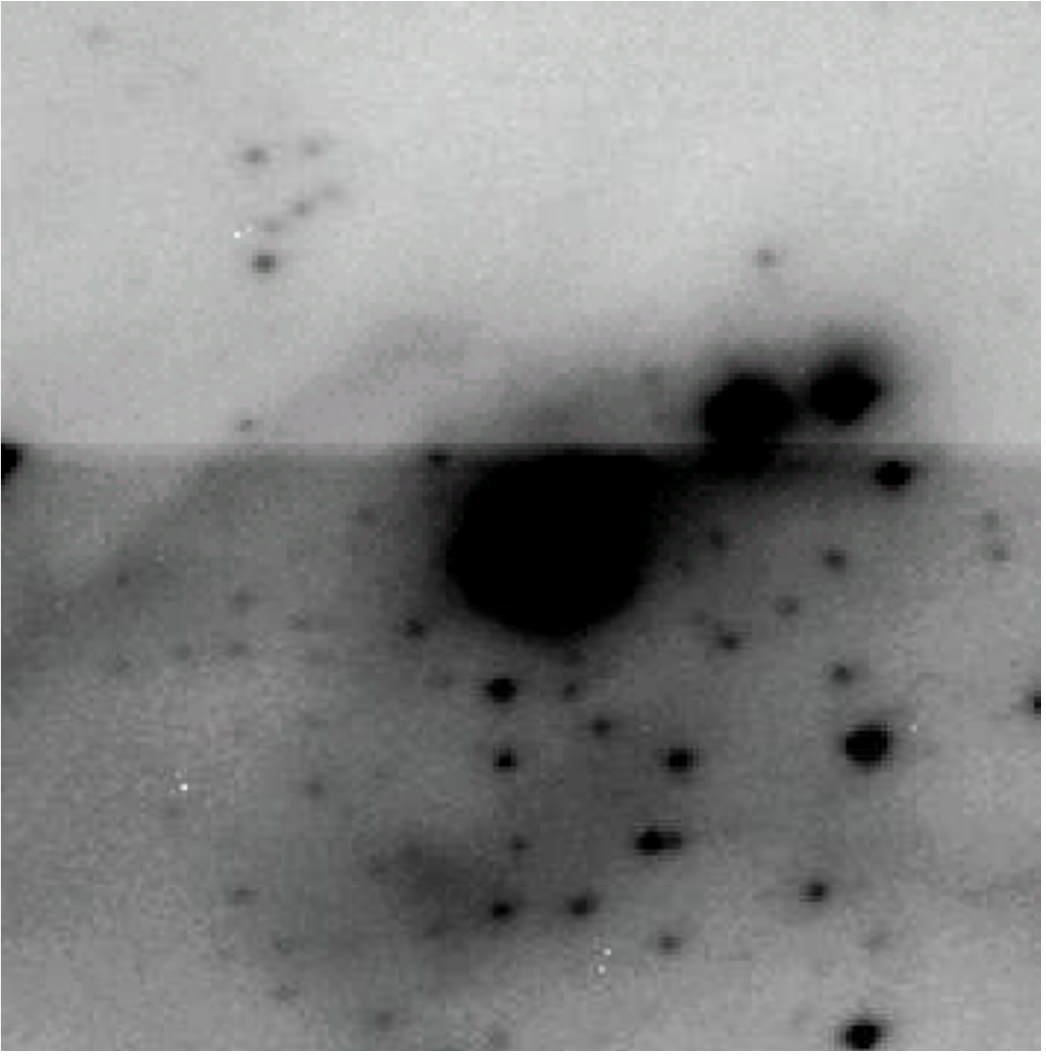}
\caption{053845.15 $-$690507.9: \h- and \ks-band.}\label{yso6}
\end{center}
\end{figure}

Classified as a `definite YSO' by GC09, their published position is
approximately 0\farcs4 north of P1222/S116/IRSN-101 \citep[][RBW98]{p93,s98}, which was
classified as O3-6~V by \citet{wb97} from ground-based spectroscopy, later
revised to O9~V(n)p by \citet{wal02} from {\em HST} observations;
5$''\times$5$''$ images of the region are shown in Figure~\ref{yso6}.
Note that P1222 was not used in the astrometric calibration of the
MAD frames and yet its recovered position agrees with Skiff's astrometry
to better than two pixels.

This source lies within the dense nebular region referred to as
`Knot~1' by \citet{wb86}, highlighted as young massive
stars just emerging from their natal cocoons and imaged at optical and
near-IR wavelengths with {\em HST} by \citet{wal99,wal02}.
The relatively bright
{\em Spitzer} magnitudes quoted by GC09 are in keeping with P1222 as
the most plausible counterpart.  Indeed, the spectral energy
distribution shown in Figure~13 of GC09 reveals increased magnitudes
bluewards of 1\,$\mu$m, as well as redwards, consistent with a massive
star with an IR excess.  From the MAD images we find \ks\,$=$\,14.31
and \h$-$\ks\,$=$\,0.13, as compared to \ks\,$=$\,14.45 and
\h$-$\ks\,$=$\,0.03 from the {\em HST}-NICMOS observations of Knot~1
by \citet{br01}.

\citet{v09} note the source as residing in a resolved H~\2 region and, on the
basis of the H$\alpha$ luminosity (uncorrected for extinction) provide a lower
bound on the inferred spectral type of the YSO of B0.5; evidently from the
classification of \citeauthor{wb97}, the star is slightly hotter.

\subsection{Spatial distribution and masses of YSOs}
It is not surprising that five of the six YSO candidates discussed
here lie to the north and west of R136 (Figure~\ref{yso_wfi}). This is
the location of the `second generation' of massive stars
\citep{wal99,wal02}, with significant molecular material still
remaining \citep[e.g.][]{w78,j98}. The sixth source ({\it 053843.52
$-$690629.0}) is spatially distinct from the regions of triggered star
formation and is the most ambiguous of the YSOs discussed here; 
with M$_K$\,$\sim$\,$-$1.4, it is also likely to be the lowest mass object.

The absolute \ks-band magnitudes of our four brightest YSO
counterparts are all bright, with M$_K$\,$\sim$\,$-$3.75-4.75,
suggesting that they are {\it bona fide} candidates for massive O- or
early B-type stars, depending on the degree of obscuration/reddening.
The most intriguing in this context is the very red source {\it
053839.69 $-$6905381} which, given the uncertainties of the
{\em Spitzer} astrometry, would certainly benefit from ground-based
spectroscopy to elucidate its nature and its relationship with P733 nearby.
\citet{r09} have recently reported a dense molecular cloud toward this
source (their IRSW-127) suggesting that it may well be a massive star 
still heavily enshrouded in its natal cocoon.

\section{Radial Luminosity Profile of R136}\label{profiles}

We now compare the radial luminosity profile for R136 from the MAD
data, constructed using a combination of integrated-light measurements and 
star counts, with the optical {\em HST} results.

\subsection{Integrated Light Profiles} 

Integrated-light measurements were used to investigate the innermost
part of the radial luminosity profile in Fields~1 and 3.  These
sum the flux within annuli around the core, where the
radius of each annulus is defined such that the total flux in each is
approximately constant.  
An {\sc idl} program was written specifically for this purpose, with standard
{\sc daophot} routines in {\sc iraf} also used as a comparison. The
{\sc idl} program calculates the integrated light by re-sampling each
image as a function of azimuth ($\phi$) and radius from the cluster
centre, to enable a smoother, more robust calculation at smaller radii
than possible with the (albeit small) sampling of the pixels -- in
practice the results are essentially the same as those obtained with
{\sc iraf}.  These measurements adopt a constant sky value, which is
defined well away from the cluster core.

To estimate the uncertainties in the integrated-light profile, each
annulus is split into azimuthal sections and the density is calculated
in each; the standard deviation of the results around each annulus
then provide an estimate of the statistical uncertainties, provided the
cluster is symmetric. However, visual inspection of the images
shows that the cluster is not symmetric, particularly at larger radii 
(see Section~\ref{azvar}), suggesting that the variations in the uncertainties
are shaped by both statistical errors and asymmetric variations of the cluster.

\begin{figure}
\begin{center}
\hspace{-0.9cm}\includegraphics[width=8.75cm]{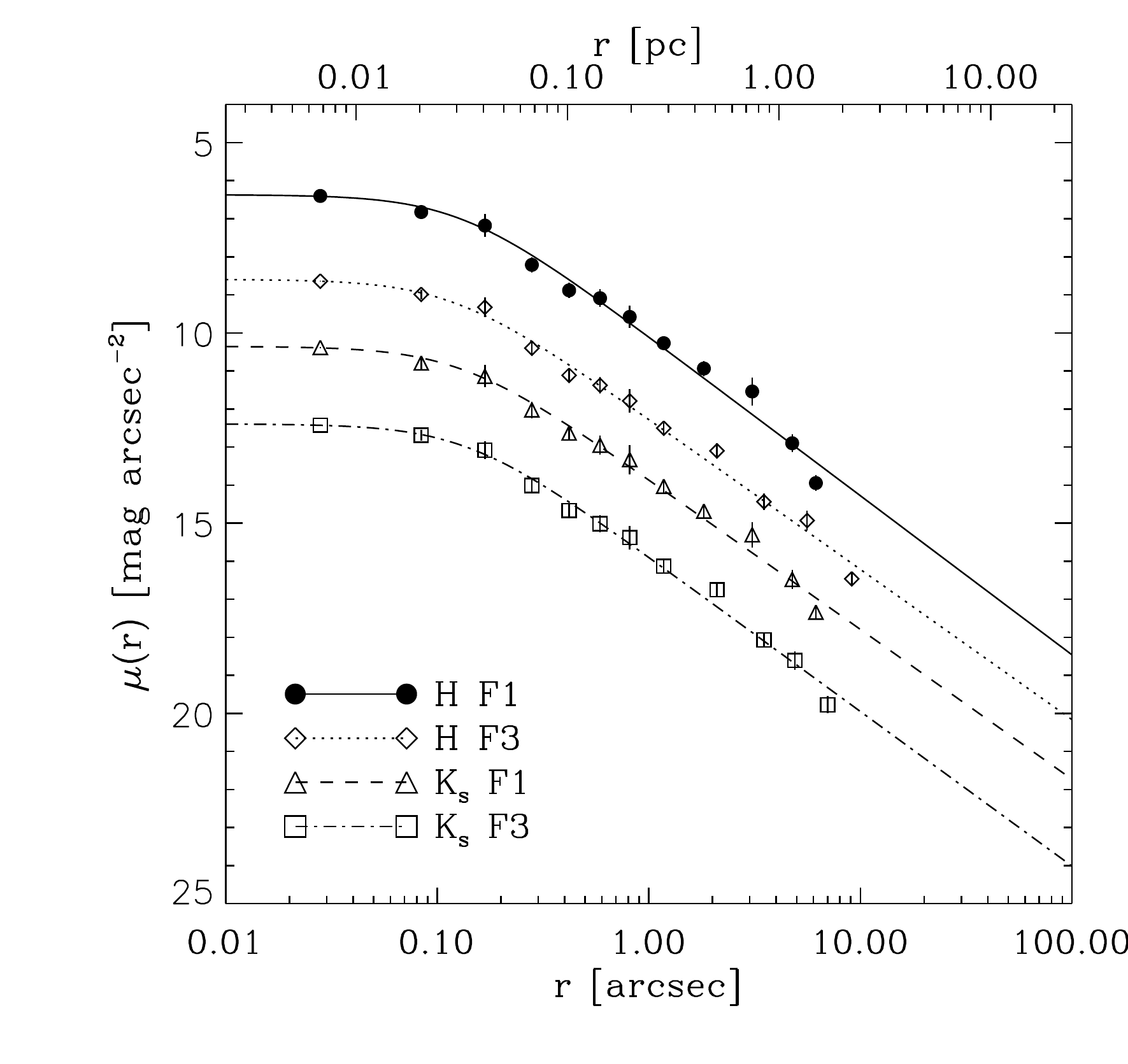}
\caption{EFF fits to the MAD integrated-light profiles, with the \ks- and \h-band (Field~3) profiles 
shifted to fainter magnitudes for clarity.}\label{eff_fits1}
\end{center}
\end{figure}

The surface brightness profiles were then fit with \citet[][EFF]{eff} cluster 
profiles of the form:
\begin{equation}
\mu(r)=\mu_0 (1+\frac{r^2}{a^2})^{-\gamma/2},
\end{equation}
where $a$ is the scale radius that defines where the flat inner part
(the core) turns into a power-law profile with an index of
$-\gamma$. The results are given in Table~\ref{eff_tab}, both in terms of angular extent on
the sky and in parsecs (adopting a distance modulus to the LMC of
18.5).  The profile fits are shown in Figure~\ref{eff_fits1}, in which the lower three profiles 
are offset (in steps of two magnitudes) for clarity.

There is hardly any signature of a core, with the surface brightness
profile fit by a power-law function over the full range of
the data.  The scale radius ($a$) obtained is effectively
$\sim$\,0.1$''$ (0.025\,pc) which is, in practice, the resolution limit
of these data; i.e. 0.1$''$ provides an upper limit to the core 
radius\footnote{Note that the core radius, $r_{\rm core}$, is usually
defined as the radius where the surface brightness profile drops to
half its central value. For an \citeauthor{eff} profile it follows
that $r_{\rm core}=a\,(2^{2/\gamma}-1)$. For the values of $\gamma$ we
find here, $r_{\rm core}\approx1.5\,a$.}.  Both the results for the core and
the slope of the power-law fit (i.e. $\gamma$\,$\sim$\,1.6) are in good
agreement with past optical {\em HST} results \citep{c92,h95} and the new
NICMOS results \citep{a09}.  

\begin{table*}
\begin{center}
\caption{Structural parameters for R136 from EFF fits to the MAD integrated-light profiles.  The italicised
results are from fits to the combined luminosity profiles (see Section~\ref{combined}).  For
convenience, results are quoted in direct observables and their physical size in pc.}\label{eff_tab}
\begin{tabular}{lcccc}
\hline
Field &  $\mu_0$ [mag arcsec$^{-2}$] &      $a$ [$''$] &          $\gamma$ &    $r_{\rm c}$ [$''$] \\
\hline
H F1 & 6.37\,$\pm$\,0.12 & 0.13\,$\pm$\,0.03 & 1.67\,$\pm$\,0.15 & 0.15\,$\pm$\,0.04\\ 
& {\em 6.36}\,$\pm$\,{\em 0.12} & {\em 0.12}\,$\pm$\,{\em 0.02} & {\em 1.57}\,$\pm$\,{\em 0.05} & {\em 0.14}\,$\pm$\,{\em 0.03} \\ 
H F3 & 6.59\,$\pm$\,0.11 & 0.12\,$\pm$\,0.03 & 1.58\,$\pm$\,0.13 & 0.14\,$\pm$\,0.03\\ 
& {\em 6.59}\,$\pm$\,{\em 0.11} & {\em 0.12}\,$\pm$\,{\em 0.02} & {\em 1.57}\,$\pm$\,{\em 0.05} & {\em 0.13}\,$\pm$\,{\em 0.02}\\ 
K F1 & 6.35\,$\pm$\,0.14 & 0.13\,$\pm$\,0.04 & 1.58\,$\pm$\,0.15 & 0.15\,$\pm$\,0.04\\ 
& {\em 6.35}\,$\pm$\,{\em 0.14} & {\em 0.12}\,$\pm$\,{\em 0.02} & {\em 1.48}\,$\pm$\,{\em 0.06} & {\em 0.15}\,$\pm$\,{\em 0.03}\\ 
K F3 & 6.39\,$\pm$\,0.10 & 0.14\,$\pm$\,0.03 & 1.63\,$\pm$\,0.14 & 0.16\,$\pm$\,0.04\\ 
& {\em 6.38}\,$\pm$\,{\em 0.11} & {\em 0.12}\,$\pm$\,{\em 0.02} & {\em 1.56}\,$\pm$\,{\em 0.06} & {\em 0.15}\,$\pm$\,{\em 0.03}\\ 
\hline
&&&&\\
\hline 
Field &      $\mu_0$ [mag pc$^{-2}$] &            $a$ [pc] &            $\gamma$ &        $r_{\rm c}$ [pc] \\
\hline
H F1 & 3.293\,$\pm$\,0.121 & 0.031\,$\pm$\,0.008 & 1.67\,$\pm$\,0.15 & 0.035\,$\pm$\,0.009\\ 
& {\em 3.282}\,$\pm$\,{\em 0.121} & {\em 0.028}\,$\pm$\,{\em 0.005} & {\em 1.58}\,$\pm$\,{\em 0.05} & {\em 0.033}\,$\pm$\,{\em 0.006} \\ 
H F3 & 3.517\,$\pm$\,0.112 & 0.028\,$\pm$\,0.007 & 1.58\,$\pm$\,0.13 & 0.034\,$\pm$\,0.008\\
& {\em 3.509}\,$\pm$\,{\em 0.112} & {\em 0.027}\,$\pm$\,{\em 0.005} & {\em 1.57}\,$\pm$\,{\em 0.06} & {\em 0.032}\,$\pm$\,{\em 0.006} \\ 
K F1 & 3.278\,$\pm$\,0.136 & 0.031\,$\pm$\,0.009 & 1.58\,$\pm$\,0.15 & 0.037\,$\pm$\,0.010\\
& {\em 3.269}\,$\pm$\,{\em 0.136} & {\em 0.029}\,$\pm$\,{\em 0.006} & {\em 1.48}\,$\pm$\,{\em 0.06} & {\em 0.035}\,$\pm$\,{\em 0.007} \\ 
K F3 & 3.316\,$\pm$\,0.105 & 0.033\,$\pm$\,0.008 & 1.63\,$\pm$\,0.14 & 0.039\,$\pm$\,0.010\\
& {\em 3.305}\,$\pm$\,{\em 0.105} & {\em 0.030}\,$\pm$\,{\em 0.006} & {\em 1.56}\,$\pm$\,{\em 0.06} & {\em 0.036}\,$\pm$\,{\em 0.007} \\ 
\hline
\end{tabular}
\end{center}
\end{table*}

\subsection{Star Counts}\label{direct}

Star counts in annuli around the core were used to calculate the
radial luminosity profile in the regions beyond $r$\,$=$\,2\farcs8.
Stars were binned such that a similar number of objects were included
per annulus, as suggested by \citet{au05}. The stars were binned as a
function of radius and magnitude and completeness terms were
calculated.  As the core is off-centre in each frame, area correction
terms were applied to correct the densities if part of a given annulus
extended past the edge of an image.

The radial completeness tests (e.g. Figure~\ref{rad_comp_Kp1}) reveal,
not unsurprisingly, that the first couple of annuli beyond
$r$\,$=$\,2\farcs8 are also strongly affected by crowding, and are
thus subject to greater uncertainties than in the outer regions. The
50$\%$ completeness level in the third radial bin was adopted as the
faint limit for construction of the profiles from the star counts,
corresponding to a faint magnitude limit in each field in the range of
18.5-18.8$^{\rm m}$, equating to a main-sequence mass of
$\sim$5\,M$_\odot$.  The innermost bins were not included in
construction of the final combined profile.

\subsubsection{Azimuthal Density Variations}\label{azvar}
To investigate the apparent asymmetry in R136 more quantitatively, we
calculated the azimuthal density variation beyond the central 2\farcs8.
An {\sc idl} program was used to split the images into angular slices,
each containing approximately equal numbers of stars (after
completeness corrections were applied).  Similar techniques have been
used recently to investigate the structure and asymmetries seen in
lower-mass Galactic clusters \citep[e.g.][]{gm05,gm09}.

Due north from the cluster core was set as $\phi$=0$^\circ$,
increasing anti-clockwise (increasing towards the east), with the
outer radius set by the distance from the cluster core to the northern
edge of Field~1.  This enables a near complete azimuthal sweep in the
combined \h-band images of Fields~1 and 3 (i.e. not the full extent of
those shown in Figure~\ref{obs}).  An iterative routine was used to
define an azimuthal slice in each image (increasing by 1$^\circ$ each
loop), which summed the star counts in that slice and corrected for
the incompleteness (down to 50\%).  Once 200 stars are obtained, the
routine determines the area covered by each slice, calculates the
density, and then moves on to define the next slice.

\begin{figure}
\begin{center}
\hspace{-0.85cm}\includegraphics[height=6cm]{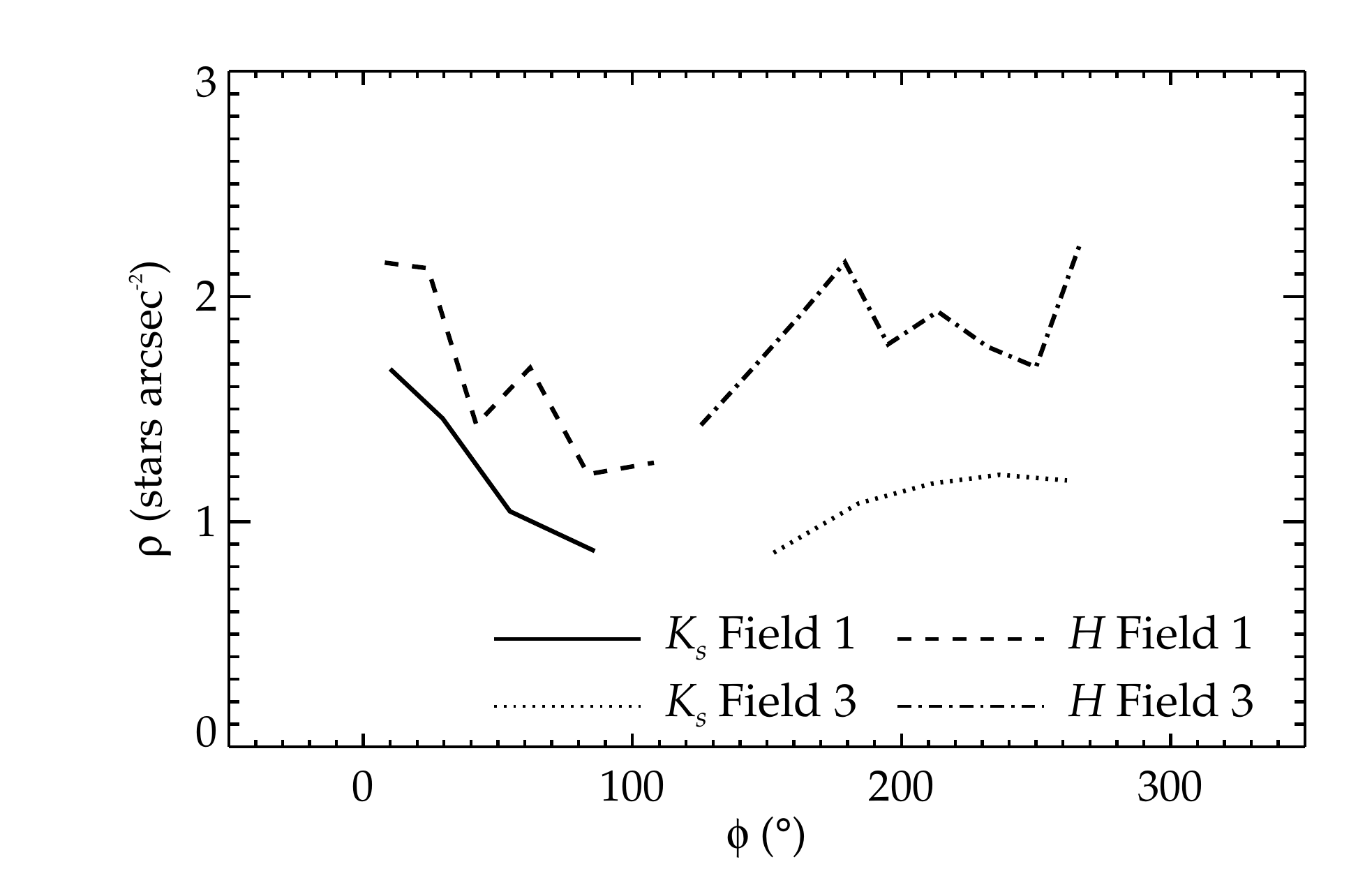}
\caption{Azimuthal density profiles for a maximum radius of 28\farcs0 from the core of R136 and
a completeness cut-off of 50\%, demonstrating that the cluster is not relaxed over a $\sim$7\,pc scale.
Due north from the cluster centre corresponds to $\phi$ = 0${^\circ}$.}\label{angular_density}

\hspace{-0.85cm}\includegraphics[height=6cm]{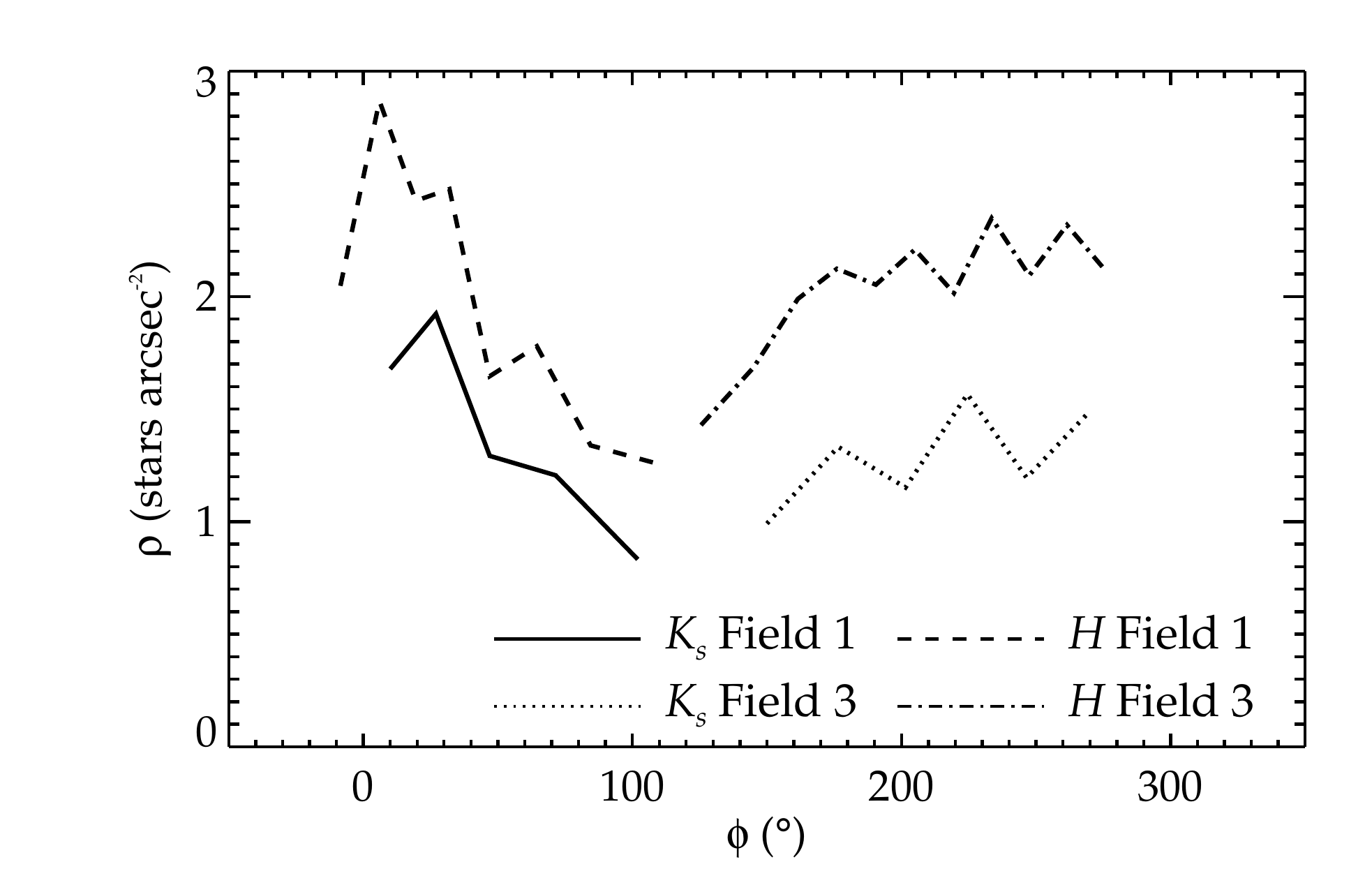}
\caption{As Figure~\ref{angular_density}, but with a completeness cut-off of 40\%.}\label{az2}
\end{center}
\end{figure}

The azimuthal variation around R136 (with $r_{\rm max}$\,=\,28\farcs0,
$\equiv$\,1,000 pixels) is clearly seen in
Figure~\ref{angular_density}, demonstrating that this region is far
from symmetric, with a local minimum around $\phi$\,$=$\,90$^\circ$.
Note that the \h-band densities are larger as a consequence of the
deeper images compared to the \ks-band observations (see
Figure~\ref{complete}).  The exact densities are sensitive to the
completeness threshold employed, so similar calculations were
undertaken for completeness cuts of 30, 40, and 60\% (with the results
for 40\% shown in Figure~\ref{az2}).  While there are variations
in the exact densities as a function of $\phi$ (partly because the
different depths yield slices which vary in size in terms of $\phi$), 
the overall trend remains the same.

Adopting a distance modulus of 18.5 to the LMC, $r$\,=\,28\farcs0
corresponds to a projected radius of 6.8\,pc.  Similar trends in the
azimuthal profiles are also seen adopting outer radii of 600 and 800
pixels (4.1 and 5.4\,pc, respectively).  At
smaller radii incompleteness effects become more significant
(cf. Figure~\ref{rad_comp_Kp1}) and it is harder to characterise the
asymmetry meaningfully.  Superficial inspection of Figure~2 from
\citet{mh98} suggests similar asymmetries in the inner region, with
luminous stars appearing to be more numerous in the eastern half of
the central 4\farcs55\,$\times$\,4\farcs55 {\em HST} image.  However,
we also note that the most luminous, massive stars will likely have a
different relaxation time to the lower-mass population.

\subsection{Combined Radial Profile}\label{combined}

The overlap region between the profiles from the two methods was used
to convert (and normalise) the star counts into surface brightness
-- a reasonable approach provided there is no evidence of mass
segregation in the cluster \citep[as claimed by][although Brandl et al. 1996
argued to the contrary]{h95}.  The average
offset in the overlap region was found from interpolation between the
two profiles, re-sampled to an evenly sampled grid of radial bins to
avoid a bias toward smaller radii (at which the bins are smaller) and
excluding the innermost bin from the star counts profile (to be less
sensitive to completeness corrections).  The offset was then used to
convert the star counts into a magnitude density, yielding a combined
profile.

\begin{figure}
\begin{center}
\hspace{-0.9cm}\includegraphics[width=7.8cm]{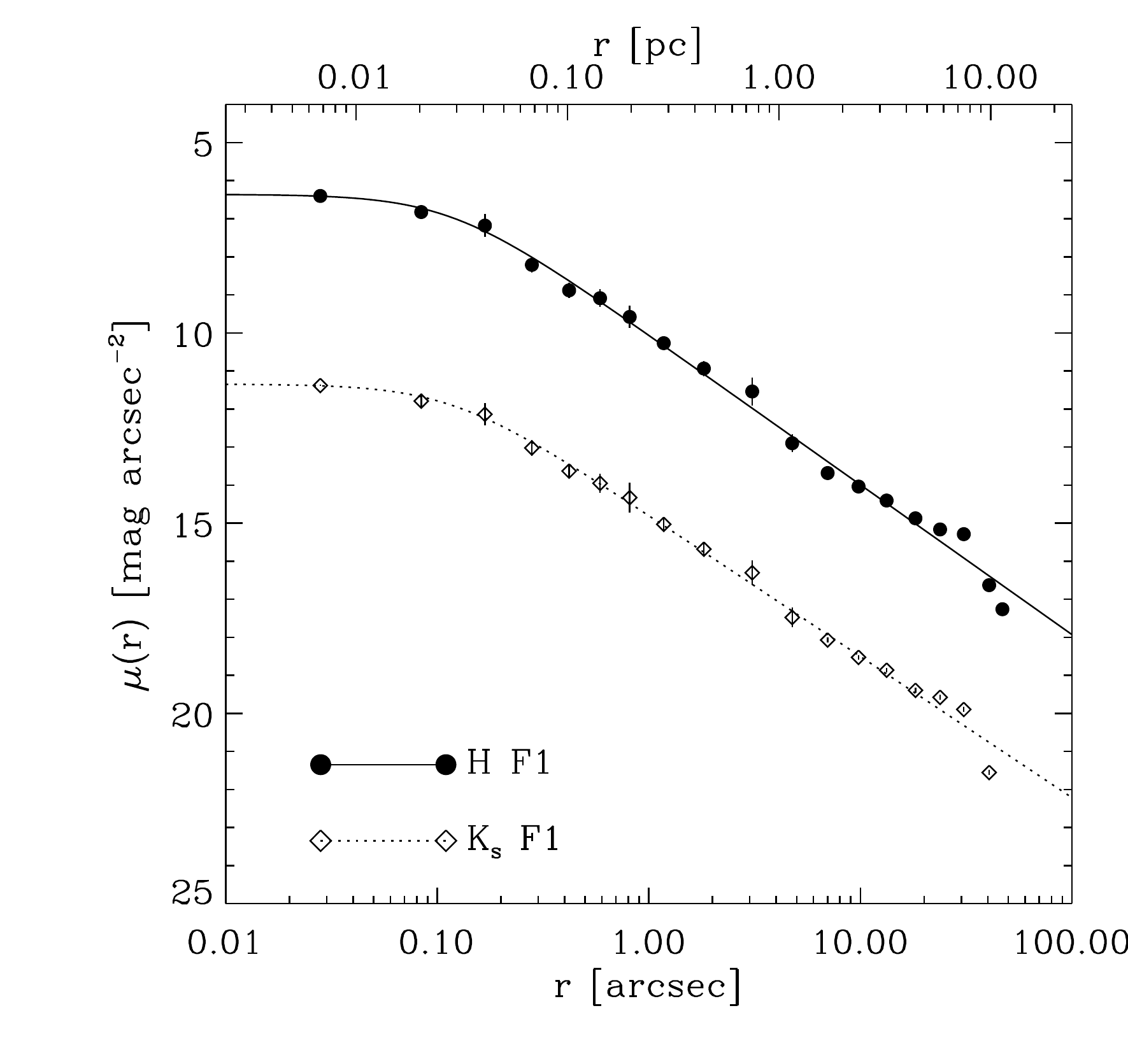}
\vspace{-0.425cm}\caption{EFF fits to the combined MAD profiles for Field~1 (with the \ks-band profile offset to fainter
magnitudes for clarity).}\label{eff_fits_f1}

\vspace{-0.25cm}\hspace{-0.9cm}\includegraphics[width=7.8cm]{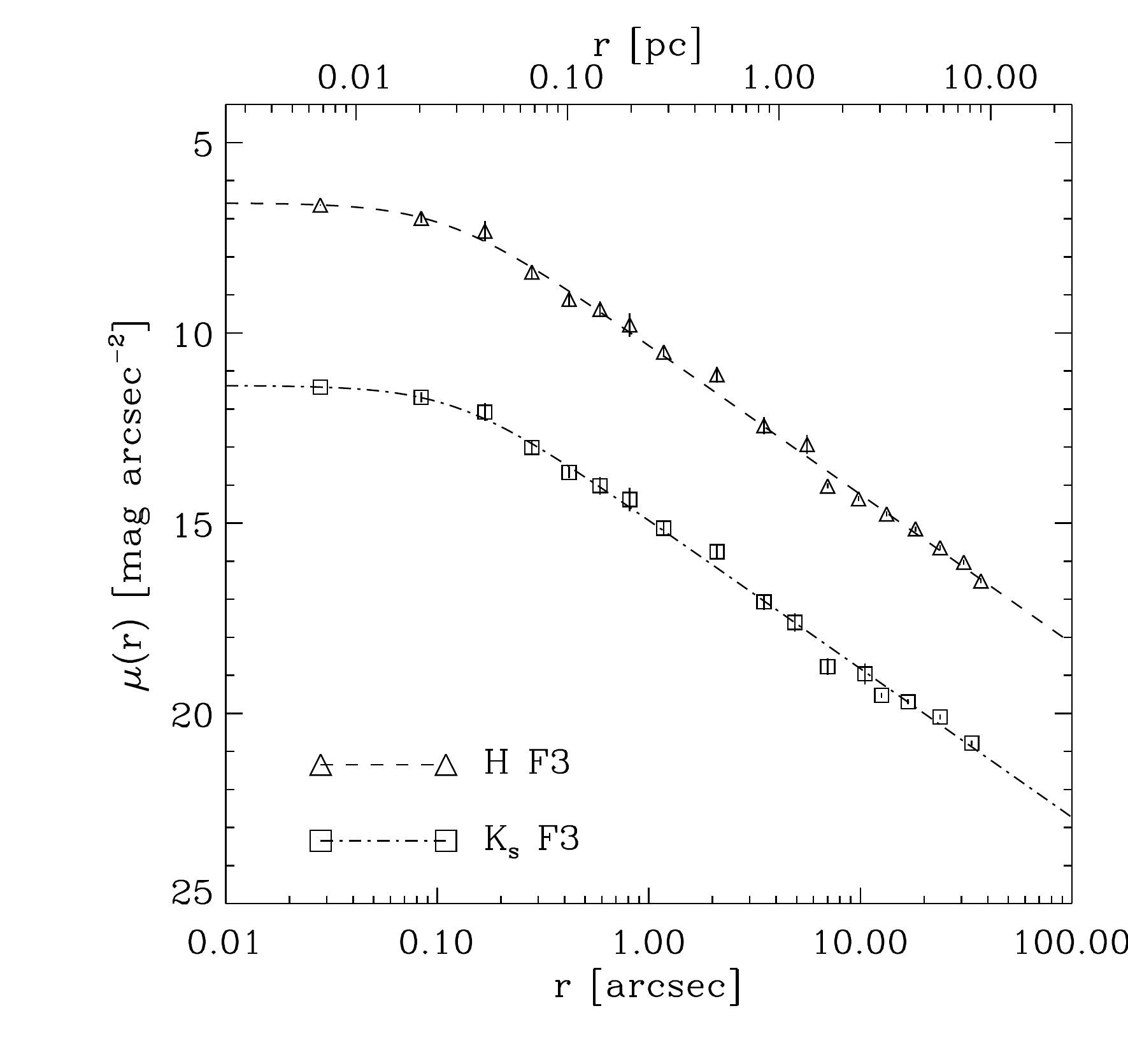}
\vspace{-0.425cm}\caption{EFF fits to the combined MAD profiles for Field~3 (with the \ks-band profile offset to fainter
magnitudes).}\label{eff_fits_f3}

\vspace{-0.25cm}\hspace{-0.9cm}\includegraphics[width=7.8cm]{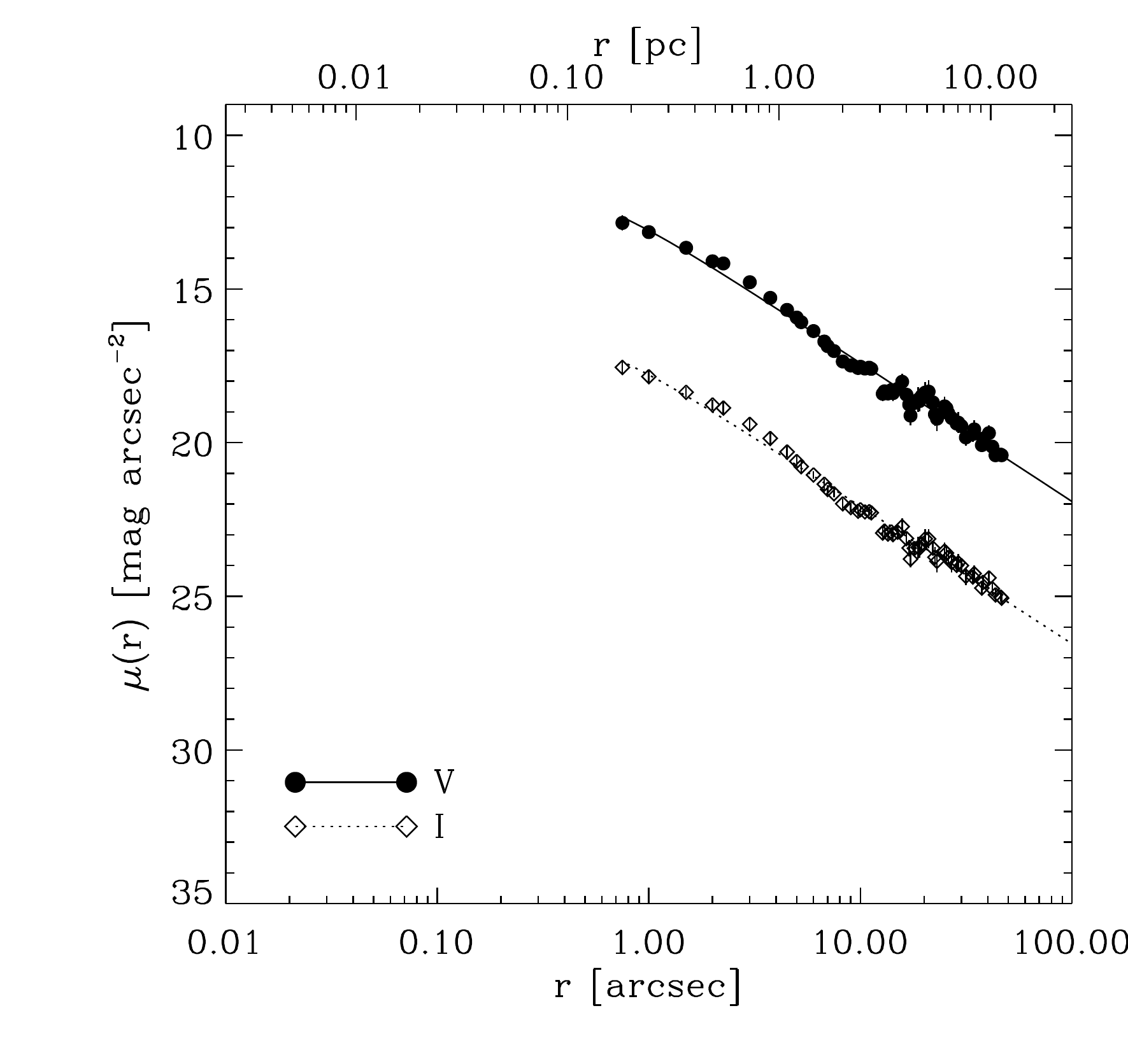}
\vspace{-0.425cm}\caption{Profile fits to the {\em HST} optical data from 
\citet{mg03}, with the $I$-band profile offset to fainter magnitudes.}\label{eff_fits3}
\end{center}
\end{figure}

EFF fits to the combined profile are shown in
Figures~\ref{eff_fits_f1} and \ref{eff_fits_f3}, with the fit
parameters given in italics in Table~\ref{eff_tab}.  Meaningful errors
on the star counts are hard to estimate due to the difficulty
of attributing uncertatinties to the completeness corrections
(i.e. blending/crowding effects) and the area corrections
(i.e. cluster asymmetries), so we adopt a conservative error of
$\pm$10\% of the density values for each bin, which are then scaled as per the
profile into a magnitude density. Note that the slopes of the fits are
nearly identical to those from analysis of the integrated-light
profile, although the results are very slightly shallower.

There are striking differences between the EFF fits to the MAD data
and the optical results from \citet{mg03}.  We also fitted EFF
profiles to the data from \citeauthor{mg03}, with it truncated to the
outer radius of the MAD data.  However, without measurements closer
to the core it is difficult to obtain a meaningful comparison for
$\mu_0$ and $a$.  Their inner data points suggest a somewhat steeper
turnover, with \citeauthor{mg03} reporting
$\gamma$\,$=$\,2.43\,$\pm$\,0.09 for their inner EFF component, but
the integrated-light MAD measurements at $r$\,$<$\,1$''$ rule out such
a slope in the near-IR.  Most relevantly, fits to the slope of the
optical data yield $\gamma$\,$=$\,1.80\,$\pm$\,0.10 in both the $V$
and $I$ bands, in reasonable agreement with the MAD data, as
illustrated in Figure~\ref{eff_fits3}\footnote{These are the original data from \citeauthor{mg03}. 
\citet{mvm05} later identified a magnitude offset of $-$0.715 (i.e. a
brighter profile) compared to Mackey \& Gilmore, who had masked out
bright stars.}.  Note the presence of the bump (or dent) noted by both \citet{m93} and
\citet{mg03}, at a radius of approximately 10$''$.  For comparison, \citet{mvm05} found
$\gamma\,=\,$2.05 from fits to the full range of the {\em HST}
data\footnote{\citet{mvm05} find $\gamma_{\rm 3D}\,=\,$3.05, which is
the slope of the 3D density profile, where $\gamma_{\rm
3D}\,=\,\gamma\,+\,$1 because of projection).}  

A value of $\gamma$\,$=$\,1.65\,$\pm$\,0.08 was also found for the
slope of the number density profile for stars brighter than
M$_V$\,$=$\,$-$5 by \citet{mds94}.  The extension of the power-law out
to $\sim$\,10\,pc makes R136 a good example of clusters with an extended
halo -- not all young clusters have such a halo, as illustrated by
\citet{mds94} through a comparison with NGC~3603, which
lacks bright stars beyond $\sim$\,1 pc. \citet{jma01} shows several
extra-galactic examples of clusters with and without an extended
halo. The origin of such halos in some clusters compared to those without
remains unclear at present.

\subsection{Discussion}\label{discussion}

\subsubsection{Structural parameters of R136}
In summary, we find slopes of $\gamma\,\le\,$2 from EFF fits to the
near-IR luminosity profile of R136.  If extrapolated to infinity such
a profile would have infinite mass, so this can not extend to very
large radii as the bulk of the mass would then be far from the core.
It remains to be seen what the origin of these very shallow profiles
is, although \citet{l04} found that shallower profiles were more
common in the younger members of their extragalactic cluster sample,
and that $\gamma$ appears to increase with age.  The development of a
core and the reduction of the central density with age is likely the
result of mass loss by stellar evolution and heating by binaries in
the centre \citep{tp00}.

Due to the very young age of R136 its density profile could be related
to the formation process of the cluster, i.e. while it is still in its earliest
stages of evolution we might see the imprint of the parent molecular
cloud.  Molecular clouds are approximately isothermal spheres in which, 
with $\gamma$\,$=$\,1 (i.e. $\gamma_{\rm 3D}$\,$=$\,2), the density
scales as $\rho$\,$\propto$\,$r^{-2}$ \cite[e.g.][]{m84}.
For example, \citet{hs06} found that the Orion Nebula Cluster (with an age of $\sim$1\,Myr) can be
well approximated by a power-law profile with $\gamma\,=\,$1.
Alternatively, the dissipationless collapse of a (gas free) stellar
system also results in a power-law density profile, and possibly a
core. The size of the core and the slope of the density profile
depends mainly on the virial temperature in the initial configuration
\citep{m84}. For a cold collapse, i.e. zero initial velocities of the
stars, \citeauthor{m84} found $\sim\,$1.5$\,<\,\gamma\,<\,\sim\,$2, with a core size of almost
zero. The `warmer' the initial conditions, the steeper the density
profile and the larger the size of the core in the final
configuration. So the observed density profile in R136 could
also be the result of a cloud collapse, with star formation occurring
in the early phases of the collapse, such that the dynamical
interactions between the stars determine the properties of the
cluster.

\citet{h95} found no evidence for mass segregation in R136.
\citet{a09} also ruled it out  for $r$\,$>$\,3\,pc ($\equiv$12\farcs4), 
finding no evidence for the flattening of the IMF below 2\,M$_\odot$ 
reported by \citet{s00}.  An early AO study of one 12\farcs8\,$\times$\,12\farcs8
quadrant of R136 by \citet{b96} sampled a similar mass range to the MAD
data, and found evidence of mass segregation above 12\,M$_\odot$. 
Whether the MAD data shows effects of the more massive stars being
preferentially in the core will be addressed elsewhere, but
if evidence for mass segregation were found in the 
inner 3\,pc, then the core radius derived here would be somewhat smaller 
than the 'primordial' radius.

\subsubsection{Structure at larger radii}

Evidence for a second component in the near-IR profiles is less
compelling than in the optical data, and we do not invoke a
second component in our EFF fits.  There is tentative evidence in
Field~1 for a `dent' at comparable radius to the
optical data (Figure~\ref{eff_fits_f1} cf. Figure~\ref{eff_fits3}), but the data for
Field~3 are fit more cleanly by a single-component profile
(Figure~\ref{eff_fits_f3}).  Unfortunately the MAD data do not lend
themselves to analysis of the integrated light out to larger radii.
While the `stitching together' of the two parts of the profile can
influence the appearance of the combined profile, exclusion of one or
two bins in the overlap region when combining them has very
little influence on the EFF results.  Indeed, now armed with knowledge of the 
minimal core, the EFF fits to the optical data are well matched at larger 
radii.  

The lack of a second component is interesting since theoretical
explanations for bumps (or halos) around young dense clusters
\citep[e.g.][]{bg06} should be independent of the wavelength in which
the profile is constructed.  The temptation is to attribute the past
(two-component) results to variable extinction in the optical images.
From the optical image of R136 (Figure~\ref{obs}) there is a notable
`void' to the north-east (running from north-west to south-east) at
$r$\,$\sim$\,10$''$, suggesting that differential extinction could be
a prime suspect.  We also note that more prosaic reasons (using
circular annuli to determine the luminosity profiles in elliptical
clusters) are now also being suggested to account for apparent breaks
in density profiles in other clusters \citep{p09}.

In this context, the asymmetric structure seen in
Figure~\ref{angular_density} must contribute strongly -- almost
certainly leading to the differences seen in the outer regions in
Figures~\ref{eff_fits_f1} and \ref{eff_fits_f3}.  Further support of
this hypothesis is provided by inspection of the original Wide Field
Planetary Camera~2 (WFPC2) images used by
\citeauthor{mg03}.  R136 was located slightly off-centre in the Planetary
Camera detector, with the Wide-Field Cameras primarily sampling the
region east of R136, i.e. Field~1 observed with MAD.
Interestingly, \citet{a09} only fit their EFF profile to the inner 2\,pc 
($\equiv$8\farcs25), noting in the caption to their Figure~12 that 
the presence of  individual bright stars introduces `the jitter in the 
surface brightness profile', hinting at the arguments that we advance here.

Such deviations from symmetry are not surprising in such a massive, intricate,
multi-population and, as yet, seemingly unrelaxed star-formation
region.  This point is reinforced by the identification of (at least)
five distinct populations in 30~Dor by \citet{wb97}, three of which
are sampled by the MAD observations: the central `Carina Phase' (rich
in early O-type stars and including R136), an older `Scorpius OB1 Phase'
of early-type supergiants throughout the central field, and a young (likely triggered)
`Orion Phase' to the north and west, mostly embedded in the gaseous
filaments visible in Figure~\ref{obs} and partially observed by Field~2.  
The stars associated with these populations in the region immediately
around R136 are neatly illustrated by Figures~3 and 4 from \citet{w86} 
-- their distribution is far from uniform, illustrative of the 
asymmetries shown in Figure~\ref{angular_density}.

The twin populations around R136 invite the question of their
formation history.  Is the Sco~OB1 Phase simply the remaining `field'
population, dispersed from a previous star-formation event?  Are the
younger, more massive Carina Phase stars formed locally, or are they
ejected members of R136?  The answer to the latter question is most
likely a combination of both.  Without comprehensive stellar and gas 
dynamics of the
different populations in 30~Dor we are limited to speculation at the
present time, but this is one of the principal motivations for the
VLT-FLAMES Tarantula Survey \citep{f2}, which has obtained multi-epoch
optical spectroscopy for over 1,000 stars and will address
these points.

The formation history of these `halo' stars goes to the heart of the
theory of infant mortality, where deviations from EFF profiles
in M82-F, NGC\,1569-A and NGC\,1705-1 were argued by \citet{bg06} to 
arise from rapid gas removal in clusters undergoing violent relaxation.
Note that the {\em HST} profile from \citet{mg03} extends
beyond a radius of 60$''$, thus also sampling the triggered
generation as well as the halo around R136, i.e. some of the excess
light reported by \citeauthor{mg03} (and, by inference, by \citeauthor{bg06})
is likely attributable to ejected stars, but that triggered star formation 
also contributes.

To echo the sentiment of \citet{wb97}, the central 2$'$ of 30~Dor
observed with MAD corresponds to a physical size of $\sim$30\,pc.  If
one projects this to a distance of 10\,Mpc, the same structures would
all be contained within $\sim$0\farcs5.  This reinforces the need to
take potential cluster asymmetries and triggered star-formation into
account when interpreting the radial luminosity profiles of distant
unresolved clusters.

\bigskip
The MAD data have provided us with a truly unique view of the central region of
30~Doradus.  In the broader context, the delivery of impressive image quality and
stability across wide fields (compared to classical AO observations)
is very encouraging in advance of future multi-laser, MCAO systems on
8-m class telescopes, and the development of wide field,
AO-corrected, imagers and spectrometers for the E-ELT.\\

\noindent {\sc acknowledgments:}
The data presented here were obtained during the MAD Science
Demonstration campaign (package name MADD-SD-EV, ID 96408).
and we are indebted to the MAD team of Paola Amico, Enrico Marchetti 
and Johann Kolb (with particular thanks to Johann for the Strehl maps).
MAC acknowledges financial support from the Science and Technology
Facilities Council (STFC).  We thank the referee, Morten Andersen, 
and Nolan Walborn and Jesus Ma\'{i}z Apell\'{a}niz for their suggestions 
and comments which improved this article.  We also thank the HAWK-I 
instrument and commissioning teams for making
the 30~Dor frames available and John Pritchard for his advice with
those images, Robert Gruendl for correspondence regarding the YSO
catalogue, Yazan Al Momany for a fateful conversation over coffee in
Santiago and subsequent advice on the individual frames method, and
Peter Stetson for making his routines available to us.

\vspace{-0.2in}

\bibliography{mad}   
\bibliographystyle{mn2e}   

\end{document}